\documentclass[twocolumn,showpacs,preprintnumbers,amsmath,amssymb]{revtex4-1}
\usepackage{graphicx}
\usepackage{dcolumn}
\usepackage{bm}
\usepackage{color}
\usepackage{physics}
\pdfoutput=1

\renewcommand{\d}{{\rm d}}
\newcommand{\e}{{\bm e}}

\newcommand{\cA}{{\bm{\mathcal A}}}
\newcommand{\B}{{\bm{B}}}
\newcommand{\cB}{{\bm{\mathcal B}}}
\newcommand{\E}{{\bm{E}}}
\newcommand{\cE}{{\bm{\mathcal E}}}
\newcommand{\J}{{\bm{J}}}
\newcommand{\cJ}{{\bm{\mathcal J}}}

\newcommand{\p}{{\bm{p}}}
\newcommand{\m}{{\bm{m}}}

\renewcommand{\j}{{\bm{j}}}

\newcommand{\q}{{\bm{q}}}
\renewcommand{\r}{{\bm{r}}}
\newcommand{\rp}{{\bm{r}'}}
\newcommand{\I}{{\mathrm{i}}}

\newcommand{\sg}{\text{sg}}
\newcommand{\tq}{{\tilde q}}

\newcommand{\smatrix}{{\underline{\sigma}}}
\newcommand{\ed}{\varepsilon}
\newcommand{\bs}{{\bar\sigma}}

\newcommand\phs{\varphi_{\text{S}}}
\newcommand\pha{\varphi_{\text{A}}}

\newcommand\hph{\widehat{\phi}}

\newcommand\arcosh{\text{arccosh}}
\newcommand\arcsinh{\text{arcsinh}}

\newcommand\jb{{\bar \jmath}}

\def\lsim{\lower.35em\hbox{$\stackrel{\textstyle<}{\textstyle\sim}$}}
\def\gsim{\lower.35em\hbox{$\stackrel{\textstyle>}{\textstyle\sim}$}}

\begin{document}

\title{Theory of plasmonic edge states in chiral bilayer systems}

\author{Dionisios Margetis$^1$ and Tobias Stauber$^{2}$}

\affiliation{$^{1}$ Institute for Physical Science and Technology, and Department of Mathematics, and Center for Scientific Computation and Mathematical Modeling, University of Maryland, College Park, Maryland 20742, USA\\
$^{2}$ Materials Science Factory,
Instituto de Ciencia de Materiales de Madrid, CSIC, E-28049 Madrid, Spain}
\date{\today}

\begin{abstract}
We analytically describe the plasmonic edge modes for an interface that involves  the twisted bilayer graphene (TBG) or other similar Moir\'e van der Waals heterostructure. For this purpose, we employ a spatially homogeneous, isotropic and frequency-dependent tensor conductivity which in principle accounts for electronic and electrostatic interlayer couplings. We predict that the edge mode dispersion relation explicitly depends on the chiral response even in the nonretarded limit, in contrast to the collective bulk plasmonic excitations in the TBG. We obtain a universal function for the dispersion of the optical edge plasmon in the paramagnetic regime. This implies a correspondence of the chiral-TBG optical plasmon to a magnetoplasmon of a single sheet, and chirality is interpreted as an effective magnetic field.  The chirality also opens up the possibility of nearly undamped acoustic modes in the paramagnetic regime. Our results may guide future near-field nanoscopy for van der Waals heterostructures. In our analysis, we retain the long-range electrostatic interaction, and apply the Wiener-Hopf method to a system of integral equations for the scalar potentials of the two layers. 
\end{abstract}

\maketitle 

\section{Introduction}
\label{sec:Intro}

The twisted bilayer graphene (TBG) has attracted immense attention due to its novel electronic phases that arise in the flat-band regime for twists near the magic angle $\theta_m\simeq 1.08^\circ$~\cite{Cao18a,Cao18b,Yankowitz19,Codecido19,Shen19,Lu19,Chen19,Xu18,Volovik18,Yuan18,Po18,Roy18,Guo18,Dodaro18,Baskaran18,Liu18,Slagle18,Peltonen18,Kennes18,Koshino18,Kang18,Isobe18,You18,Wu18b,Zhang18,Gonzalez19,Ochi18,Thomson18,Carr18,Guinea18,Zou18,Gonzalez20,GonzalezSB20,StauberC20}. 
 Furthermore, the plasmonic properties of the TBG indicate several surprising features not present in the usual two-dimensional (2D) systems such as the monolayer graphene~\cite{Fei12,Chen12,Koppens11,Grigorenko12,Stauber14,Peres16,Basov16,Low17}. Apart from a modified gate dependence~\cite{Hu17}, there is, e.g., the possibility of exciting collective charge oscillations at the neutrality point that are only composed of charge densities induced by interband transitions~\cite{Stauber16,Hesp19}. This possibility is due to the localization of the electronic wave function for twist angles $\theta\lesssim2^\circ$ that provides the restoring force needed to sustain the charged in-phase oscillations of the electron and hole densities. In addition, for minimal twist angles, the lattice relaxation-induced domain walls between the two equivalent Bernal-stacked configurations may act as a periodic potential for plasmons, opening up the prospect of photonic crystals for nanoscale light~\cite{Sunku18}. Novel chiral plasmons consisting of topologically protected electronic domain-wall states are also predicted if the chemical potential lies inside the energy gap~\cite{Brey20}. Lastly, plasmons in flat bands are extremely long-lived since they are unlikely to couple and decay into the particle-hole continuum~\cite{Levitov19,Khaliji20} with non-reciprocal dispersion~\cite{Papaj20}.

The above features concern the flat-band regime or lower energies. Nevertheless, Moir\'e van der Waals heterostructures also display an inherent handedness independent of the twist angle, as one can rotate the top layer to the right or to the left. This structural chirality is passed onto the electronic properties. Consequently, optical dichroism is observed when the bilayer system is coupled to circularly polarized light~\cite{Kim16,Suarez17}. In fact, plasmonic properties are inherently chiral~\cite{Stauber18,Stauber18b} due to the quantum mechanical interlayer coupling. The associated electromagnetic near-fields may pave the way to promoting chiral chemistry~\cite{Stauber20NL}. However, the plasmonic dispersion relation only depends on the chiral structure in the retarded regime. Thus, the chiral effect is small~\cite{Lin20}.

In this paper, we analytically investigate how the chirality of the bilayer system affects the dispersion relation of edge modes in the quasi-electrostatic limit. We use a minimal model with an effective isotropic, spatially homogeneous and frequency dependent conductivity tensor~\cite{Stauber18}, represented by a $4\times 4$ matrix, which can in principle capture electronic and electrostatic interlayer couplings of the TBG. Our analysis explicitly shows how chirality couples the optical and acoustic edge modes and thus modifies their dispersion in the nonretarded limit. We obtain a universal function that describes the dispersion of optical edge plasmons when the susceptibility to an in-plane magnetic field is paramagnetic. This regime occurs for chemical potentials close to the neutrality point~\cite{Stauber18}, when the counterflow Drude weight becomes negative. \color{black} We also point out the possible existence of nearly undamped acoustic edge plasmons with linear dispersion for strong enough chirality. An assumption in our study is that the sound velocity is larger than the Fermi velocity, which enables us to use a spatially local conductivity in Maxwell's equations.  

Regarding previous works on the TBG, only bulk plasmonic excitations have been considered so far; see, e.g.,~\cite{Fei15b,Stauber16,StauberPRB12,Brey20,Stauber18,Stauber20NL,Lin20,Kuang21}. On the other hand, it is well known that at an interface collective plasmonic modes may arise with an electromagnetic field that is localized near edges. These modes have been discussed in the context of magnetoplasmons supported by a homogeneous medium~\cite{Fetter85,Volkov86,Volkov88,Wang11}; for related studies, see~\cite{Goncalves17,YouGoncalves19,Margetis20,DM20,CohenGoldstein18}. Such edge modes have recently been detected in graphene by infrared nano-imaging, i.e., scattering-type scanning near-field optical microscopy (s-SNOM)~\cite{Fei15,Nikitin16}. This type of mode usually exists for a broad class of interfaces~\cite{Stauber2DM19} and can further be launched in one direction by an appropriately polarized dipole, thus opening up unprecedented technological possibilities. In fact, the area of topological plasmonics is a rapidly emergent subfield of nanophotonics~\cite{Novotny12} based on 2D materials~\cite{Reserbat-Plantey21}. Notably, in a periodically patterned system, and in the presence of an out-of-plane magnetic field, band-structure theory yields a nontrivial Chern number, making plasmons topologically protected wave modes that can travel around obstacles~\cite{Jin17,Pan17}. Hence, it would be technologically desirable to explore the possible existence and control of plasmonic edge modes in the \emph{chiral} TBG, adding a knob for tuning their dispersion by chirality. These modes can be observable by, e.g., scattering scanning near-field microscopy which is a powerful technique in the context of both the monolayer graphene and TBG~\cite{Sunku20,Sunku21,Hesp21}. 

We emphasize that bulk plasmons in twisted heterostructures are intrinsically chiral~\cite{Stauber18}. However, in the retarded frequency regime, the plasmon dispersion relation is not altered and chirality only manifests itself in the near-field~\cite{Stauber20NL}. Contrary to this situation, here we will show that the coupling of the longitudinal and transverse channels occurs at the edge similarly to the case of Berry plasmons that are chiral due to a non-trivial Berry curvature~\cite{Song16,Kumar16}. This coupling can also be achieved by scattering from impurities. 

In this work, we investigate the dispersion of plasmonic edge modes that emerge at the interface of a chiral bilayer sample with an unbounded dielectric medium. We formulate a system of integral equations for the scalar potentials in two semi-infinite layers. By a linear transformation, the field equations are coupled only at the edge through chirality. We apply a variant of the Wiener-Hopf method~\cite{Masujima-book} to solve this system exactly. Our analytical predictions address both the cases of the neutrality point and finite doping. In the latter case, we show that the frequency of the optical edge plasmon is blue-shifted by chirality. In addition, the optical mode of the non-magnetic chiral TBG can exhibit a dispersion similar to that of an edge magnetoplasmon in a single sheet. In this correspondence, chirality plays the role of an effective magnetic field that can become of the order of hundreds of Tesla. Our model for the conductivity tensor can include an out-of-plane magnetic field, and thus break time-reversal symmetry and yield non-reciprocal edge modes. Other extensions, e.g., the joint effect of anisotropy and chirality, lie beyond the scope of this paper and will be addressed elsewhere. 

The remainder of the paper is organized as follows. In Sec.~\ref{sec:Formulation}, we formulate integral equations for the scalar potential in the TBG via the quasi-electrostatic approach. Section~\ref{sec:Dispersion} focuses on the derivation of the edge mode dispersion relation by the Wiener-Hopf method. In Sec.~\ref{sec:Numerics-iso}, we discuss the effect of chirality via approximations of the dispersion relation. Section~\ref{sec:Conclusion} concludes the paper. The appendices provide requisite technical derivations.

{\em Notation.} Boldface symbols such as $\E$ denote vectors. The symbol $\e_\ell$ is the unit Cartesian vector in the positive $\ell$-direction ($\ell=x,\,y,\,z$). Underlined symbols, e.g., $\smatrix$, denote square matrices. The first (second) partial derivative of $f$ with respect to $\ell$ is  $\partial_\ell f$ ($\partial_\ell^2 f$). The symbol $f(a^{\pm})$ indicates the limit of $f(x)$ as $x$ approaches $a$ from above ($+$) or below ($-$). We write $f=\mathcal O(g)$ if $|f/g|$ is bounded in a prescribed limit. The hat on top of a symbol, e.g., $\widehat{f}(\xi)$, denotes the Fourier transform of a function, e.g., $f(x)$, with respect to $x$; $\xi$ is the wave number (Fourier variable). The $+$ or $-$ \emph{subscript} in the symbol $Q_\pm(\xi)$ (not to be confused with the frequency $\omega_\pm$ of an optical or acoustic mode), where $\xi$ is a complex variable, implies that $Q_\pm(\xi)$ is analytic for $\pm \Im\xi>0$. The time-harmonic fields have the temporal dependence $e^{-\I\omega t}$ where $\omega$ is the angular frequency ($\I^2=-1$).


\section{Field equations in isotropic bilayer system}
\label{sec:Formulation}

In this section, we formulate the field equations for the edge states of an isotropic bilayer system in the non-retarded limit. In other words, we assume that the wave number, $q$, of an edge state satisfies $|q|\gg \omega/c$, where $c$ is the light speed in vacuum, applying the quasi-electrostatic approximation. This theory forms an extension of previous works for isotropic monolayer systems~\cite{Volkov88,Margetis20}. For a general derivation of the underlying electric-field integral equations with retardation effects in the TBG, see Appendix~\ref{app:int-eq}. An extension of the quasi-electrostatic theory to include anisotropy of the bilayer system will be discussed elsewhere~\cite{Margetis21}.

\begin{figure}[h]
\includegraphics[scale=0.35,trim=0in 0.8in 0in 0.4in]{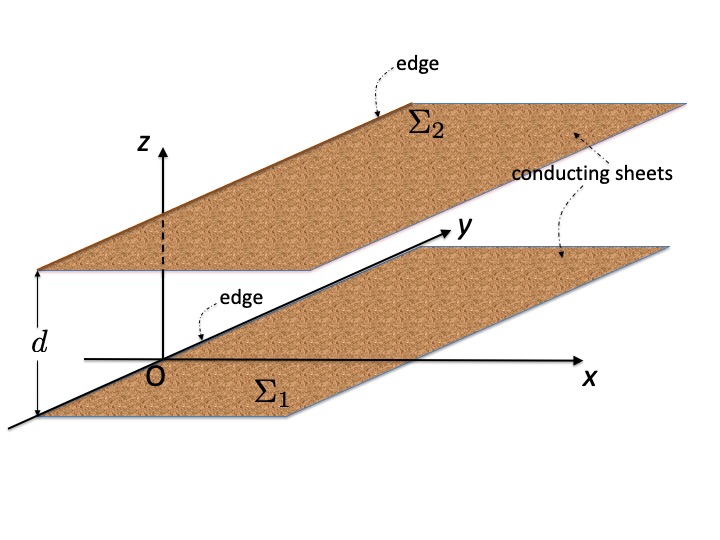}
\centering{}\caption{Geometry of the TBG system.
Two semi-infinite, flat conducting sheets, $\Sigma_1$ and $\Sigma_2$, are parallel to each other at a distance equal to $d$. Sheet $\Sigma_1$ lies in the $xy$-plane ($z=0$) for $x>0$. Layer $\Sigma_2$ lies in the plane $z=d$ for $x>0$. The layers are immersed into a homogeneous unbounded medium of dielectric permittivity $\ed$ and  magnetic permeability $\mu$.}
\label{fig:Geometry}
\end{figure}

The geometry is depicted in Fig.~\ref{fig:Geometry}. This consists of two  flat  sheets, $\Sigma_1$ and $\Sigma_2$, that lie parallel to each other at distance $d$ and have coplanar edges. The layers occupy the half planes at $z=0$ and $z=d$ in regions of positive $x$ coordinate; thus, the sheets have edges parallel to the $y$-axis. The ambient medium is homogeneous with (scalar) dielectric permittivity $\ed$ and  magnetic permeability $\mu$. Losses in this medium are included via a complex-valued $\ed$. Note that the geometry is translation invariant in $y$.

Regarding edge states, we assume that there is no externally applied source and all fields have the $e^{\I q y}$ dependence on $y$, where $\omega$ must be determined as a function of the wave number $q$ (or vice versa). From now on, we suppress the (exponential) $y$-dependence of fields. 

Let $\varphi(x,z)$ denote the electrostatic potential generated everywhere by the electron surface charge densities excited on the two layers. By using the Green function or propagator, $\mathcal G(x,z)$, of the 2D Poisson equation, we have
\begin{equation}\label{eq:phi-G-rho}
\varphi(x,z)=\ed^{-1} \iint\displaylimits_{-\infty}^{\infty}\d x' \d z'\,\mathcal G(x-x', z-z')\,\rho(x',z')	
\end{equation}
where $\rho(x,z)$ is the volume charge density, viz.,
\begin{equation*}
\rho(x,z)=\varrho_1(x)\,\delta(z)+\varrho_2(x)\,\delta(z-d)~.	
\end{equation*}
Here, $\varrho_j$ is the surface charge density on sheet $\Sigma_j$ ($j=1,\,2$) and $\delta(z)$ is Dirac's delta function; $\varrho_j(x)=0$ if $x<0$. The 2D propagator is~\cite{Volkov88,Margetis20}
\begin{equation}\label{eq:G-def}
\mathcal G(x,z)=\frac{1}{2\pi} K_0(\tq\sqrt{x^2+z^2})~,\quad \tq= q\,\sg(q)~, 	
\end{equation}
where $K_0$ is the third-kind modified Bessel function of zeroth order, and the `complex signum' function is  $\sg(q)=\pm 1$ if $\pm\Re q>0$. Thus, $\tq=q\,\text{sgn}(q)=|q|$ if $q$ is real. We stress that the propagator $\mathcal G(x,z)$ incorporates the long-range electrostatic interaction, in contrast to the kernel approximation by an exponential in~\cite{Fetter85}; thus, $\mathcal G(x,z)=\mathcal O\big(\ln(\sqrt{x^2+z^2})\big)$ near the origin. 

As an alternative to a singular propagator, we will also discuss how the edge mode is affected by the use of a regularized propagator.  This replacement amounts to the broadening of the material edge in the horizontal ($x$-) or vertical ($z$-) direction. 
In principle, the regularization procedure is not uniquely defined. We make a choice that preserves the character of the kernel as the Green function of the 2D Helmholtz equation. This choice has some advantages, e.g., the potential $\varphi(x,z)$ satisfies the 2D Helmholtz equation. 
In this vein, replace $\mathcal G(x,z)$ by 
\begin{equation*}
\mathcal G_{\text{reg}}(x,z)=\frac{1}{2\pi}K_0\big(\tq\sqrt{x^2+z^2+b^2}\big)~,\quad b>0~.
\end{equation*}
The length $b$ is of the order of or larger than $d$ ($|qb|\ll 1$). This $b$ should be chosen separately for ``symmetric'' and ``antisymmetric'' edge states (see Sec.~\ref{subsec:transf-pot}). \color{black}

The potential $\varphi(x,z)$ is continuous and the densities $\varrho_{j}(x)$ must be integrable in order to yield finite charges. These densities satisfy the continuity equation
\begin{equation*}
	-\I\omega\varrho_j(x)+\nabla_{\Sigma}\cdot \cJ_j(x)=0~,\quad -\infty<x<\infty~,
\end{equation*}
where $\nabla_{\Sigma}=(\partial_x, \I q)$ and $\cJ_j$ is the 2-component surface current density on $\Sigma_j$; $\cJ_j(x)=0$ if $x<0$.  

We invoke Ohm's constitutive law which relates the surface current densities, $\cJ_j$, of the sheets to the electric field. We assume that this law is local and homogeneous, and involves only the tangential electric field; thus,
\begin{subequations}\label{eqs:J-sigma-E}
\begin{equation}\label{eq:J-sigma}
\begin{pmatrix}
\cJ_1(x) \\
\cJ_2(x)	
\end{pmatrix}
=
\smatrix 
\begin{pmatrix}
\cE_\parallel^1(x) \\
\cE_\parallel^2(x)	
\end{pmatrix}\ x>0~;\ 
\smatrix=
\begin{pmatrix}
\smatrix_{11} & \smatrix_{12} \\
\smatrix_{21} & \smatrix_{22}	
\end{pmatrix}~.
\end{equation}
In the above, $\cE_\parallel^j(x)$ is the electric field parallel to the $xy$-plane in sheet $\Sigma_j$, at $z=0$ for $j=1$ and $z=d$ for $j=2$. The parameter $\smatrix$ is the $4\times 4$ conductivity matrix which captures the electrostatic and electronic couplings of the layers. For a minimal model that expresses isotropy with an out-of-plane magnetic field, which is perpendicular to the sheets, we define the $2\times 2$ matrices 
\begin{align}
 \smatrix_{11}&=\smatrix_{22}=
 \begin{pmatrix}
 	\sigma_0 & \sigma_B \\
 	-\sigma_B & \sigma_0
 \end{pmatrix}~,\label{eq:sigma-model1} \\
 \smatrix_{12}&=
 \begin{pmatrix}
  \sigma_1 & \sigma_2 +\sigma_B'\\
   -\sigma_2-\sigma_B' & \sigma_1 	
 \end{pmatrix}~,\label{eq:sigma-model2} \\
\smatrix_{21}&=
 \begin{pmatrix}
  \sigma_1 & -\sigma_2+\sigma_B' \\
   \sigma_2-\sigma_B'& \sigma_1 	
 \end{pmatrix}~.\label{eq:sigma-model3}
\end{align}
\end{subequations}
The  matrix elements $\sigma_0$, $\sigma_1$, and $\sigma_2$ are spatially constant and depend on material and geometry parameters such as the doping of graphene sheets or the twist angle and also the interlayer spacing, $d$, as well as the frequency, $\omega$.  Note that \emph{the parameter $\sigma_2$ expresses the chirality of the system}.  The matrix elements $\sigma_B$ and $\sigma_B'$ may arise from a magnetic field perpendicular to the sheets~\cite{Fetter85,Volkov88}. 

Next, we discuss the relation of matrix elements of $\smatrix_{ij}$ to in-plane dipoles in some generality. The electric in-plane dipole, $\p_\parallel$, is related to the sum of two-sheet currents, $\j_1$ and $\j_2$, viz., $-\I\omega\p_\parallel=\j_1+\j_2$; whereas the magnetic in-plane dipole, $\m_\parallel$, is given by the difference of the two-sheet currents, $\m_\parallel=d\e_z\times(\j_2-\j_1)/2$. The constituent equations read
%
\begin{align*}
\p_\parallel&=-2\frac{\sigma_0+\sigma_1}{i\omega}\E_\parallel-2\frac{\sigma_B+\sigma_B'}{i\omega}\e_z\times\E_\parallel+d\sigma_2\B_\parallel~,\\
\m_\parallel&=d\sigma_2\E_\parallel+i\omega\frac{d^2}{2}\left[(\sigma_0-\sigma_1) +(\sigma_B-\sigma_B')\e_z\times\right]\B_\parallel~,
\end{align*}
%
where $(\E_\parallel, \B_\parallel)$ is the in-plane electromagnetic field. \color{black}
The model is invariant under rotation, and the Onsager relations are fulfilled if $\sigma_B^{(\prime)}$ changes sign according to the magnetic field component $B$ perpendicular to the sheets, i.e., $\sigma_B^{(\prime)}=\text{sgn}(B)\sigma_{|B|}^{(\prime)}$. In our notation for $\sigma_B^{(\prime)}$ we use the vertical ($z$-) component $B=B_\perp$, not to be confused with the dynamic in-plane magnetic field $\B_\parallel$. In the absence of an out-of-plane magnetic field and for $\sigma_2=0$, the system resembles an ordinary double-layer system (without chirality). Let us emphasize that a finite chiral coupling, if $\sigma_2\neq0$, endows the system with chiral plasmons even without breaking time-reversal symmetry (if $B=0$)~\cite{Stauber18}.

\begin{figure}[h]
\includegraphics[scale=0.35,trim=0in 0.3in 0in 0.4in]{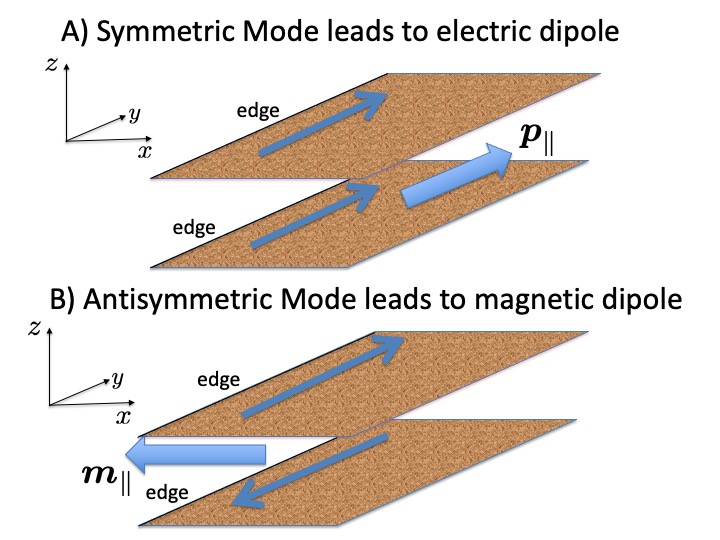}
\centering{}\caption{Schematic of currents giving rise to electric dipole $\boldsymbol p_\parallel$ (A) and magnetic dipole $\boldsymbol m_\parallel$ (B) of edge modes in the TBG.}
\label{fig:Dipoles}
\end{figure}

In our model, $\sigma_B$ denotes the in-plane Hall response and $\sigma_B'$ resembles the response function on layer 1 due to the transverse drag of a current in layer 2. A simple model based on the equations of motion yields that these functions are proportional to the in-plane and drag conductivities, $\sigma_0$ and $\sigma_1$. For weak enough out-of-plane magnetic field, we have $\sigma_B=-i(\omega_c/\omega)\sigma_0$ and $\sigma_B'=-i(\omega_c/\omega)\sigma_1$, where $\omega_c=eB/m$ is the cyclotron frequency. Hence, the constituent equations become 
\begin{subequations}\label{eqs:ConEqBeta}
\begin{align}
\p_\parallel&=-2\frac{\sigma_0+\sigma_1}{i\omega}\left(\E_\parallel-i\frac{\omega_c}{\omega}\e_z\times\E_\parallel\right)+d\sigma_2\B_\parallel~,\\
\m_\parallel&=d\sigma_2\E_\parallel+\frac{i\omega d^2}{2}(\sigma_0-\sigma_1)\left(1-i\frac{\omega_c}{\omega}\e_z\times\right)\B_\parallel~.
\end{align}
\end{subequations}
Let us stress that we will in principle treat $\sigma_B$ and $\sigma_B'$ as independent parameters, unless stated otherwise.
A schematic of the currents and the two types of dipoles for the edge modes in the TBG is shown in Fig.~\ref{fig:Dipoles}.

There are two obvious extensions of this model. First, the symmetry of the two layers can be broken under different conductivities and Hall response that would preserve the rotational invariance. The second extension is to assume a birefringent system with different in-plane conductivities in the $x$- and $y$-directions which would break rotational symmetry. The latter extension can be carried out and will be discussed elsewhere~\cite{Margetis21}. The former extension will be the subject of future work. 

\subsection{System of integral equations}
\label{subsec:int-eqs-isot}

Next, we derive integral equations for $\varphi_1(x)=\varphi(x,0)$ and $\varphi_2(x)=\varphi(x,d)$, which take into account the electrostatic and electronic interlayer couplings. The starting point is Eq.~\eqref{eq:phi-G-rho} for the potential $\varphi(x,z)$ in terms of the volume charge density, $\rho(x,z)$. We express this $\rho$ in terms of surface charge densities on the sheets; invoke  the continuity equation on each layer; use Ohm's law~\eqref{eqs:J-sigma-E} for the surface current densities; and apply the quasi-electrostatic approximation in the form 
\begin{equation*}
\cE_\parallel^j(x)=-\nabla_{\Sigma}\varphi_j(x)=-(\partial_x, \I q)\varphi_j(x)\quad (j=1,\,2)~.
\end{equation*}
Here, $\nabla_\Sigma=(\partial_x, \I q)$ denotes the gradient in the $xy$-plane.
The potential $\varphi(x,z)$ arises from the surface charge induced on both sheets which  depends on $\varphi$ by Ohm's law. A similar procedure can be found in~\cite{Volkov88} for deriving an integral equation for magnetoplasmons; see also~\cite{Fetter85}.

Next, we enforce the condition of vanishing surface current densities normal to each edge, $\e_x\cdot \cJ_j(x)=0$ at $x=0^+$ ($j=1,\,2$). This condition is implied by the absence of any charge accumulation at each edge, and naturally comes from the electric field integral equations in the quasi-electrostatic limit; see Appendices~\ref{app:int-eq} and~\ref{app:qs}.  Hence, after an integration by parts in Eq.~\eqref{eq:phi-G-rho}, we write 
\begin{align*}
\varphi(x,z)=&\frac{1}{\I\omega\ed}(\partial_x, \I q)\cdot \left\{\int_0^\infty \d x'\,\left[\mathcal G(x-x',z) \cJ_1(x')\right.\right. \notag\\
&\mbox{}\left.\left. \quad +\mathcal G(x-x',z-d) \,\cJ_2(x')\right]\right\}~,\ \mbox{all}\ (x,z)~,	
\end{align*}
where 
\begin{equation*}
\cJ_j(x)=-\smatrix_{j1}
\begin{pmatrix}
\partial_x \\
 \I q
\end{pmatrix} 
\varphi_1(x)-\smatrix_{j2}
\begin{pmatrix}
\partial_x \\
 \I q
\end{pmatrix} 
\varphi_2(x)~.
\end{equation*}
The desired integral equations result from applying integration by parts once more, and setting $z=0,\,d$ for $\varphi(x,z)$. Thus, we obtain the following system for the two layers labeled by $j=1,2$ (with $\jb=2,1$, respectively):
\begin{widetext}
\begin{align}\label{eq:phi1-int}
\varphi_j(x)&=\frac{\I\omega\mu}{k_0^2}(\partial_{x}^2-q^2)\left\{\int_0^\infty \d x'\,[\sigma_0\mathfrak K_\parallel(x-x')
+\sigma_1\mathfrak K_\perp(x-x')]\,\varphi_j(x')	
 +\int_0^\infty \d x'\,[\sigma_1 \mathfrak K_\parallel(x-x')+\sigma_0 \mathfrak K_\perp(x-x')] \varphi_{\jb}(x')\right\}\notag\\
&-\frac{\I\omega\mu}{k_0^2}\left\{[(\sigma_0\partial_x-\I q\sigma_B) \mathfrak K_\parallel(x)+(\sigma_1\partial_x-\I q\sigma_B')\mathfrak K_\perp(x)]\varphi_j(0^+)
 +[(\sigma_0\partial_x-\I q\sigma_B)\mathfrak K_\perp(x)+(\sigma_1 \partial_x-\I q\sigma_B') \mathfrak K_\parallel(x)]\varphi_{\jb}(0^+)\right.\notag\\
 &\left.+(-1)^j\I q\sigma_2[-\mathfrak K_\perp(x)\varphi_j(0^+)+\mathfrak K_\parallel(x)\varphi_{\jb}(0^+)]\right\},\qquad k_0^2=\omega^2\mu\varepsilon~,
\end{align}
\end{widetext}
for all $x$. The kernels $\mathfrak K_\parallel$ and $\mathfrak K_\perp$  express the propagator $\mathcal G(x,z)$ at $z=0,\, d$, viz.,  
\begin{equation}\label{eq:kernel-defs}
\mathfrak K_\parallel(x)=\mathcal G(x,0)~,\ \mathfrak K_\perp(x)=\mathcal G(x,d)~,	
\end{equation}
where $\mathcal G(x,z)$ is given by Eq.~\eqref{eq:G-def}. 

The problem of the edge modes can be stated as follows: For given wave numbers $q$, we need to determine the frequencies $\omega(q)$ so that Eq.~\eqref{eq:phi1-int} has nontrivial continuous and integrable  solutions 
$(\varphi_1(x),\varphi_2(x))$ for all $x$~\cite{Note}. This integrability here implies decay of $\varphi(x,0)$ away from the edge, and  localization of the mode. Alternatively, for given $\omega$ we should find $q(\omega)$. The continuity of the scalar potential at each edge is crucial in establishing the dispersion relation, by analogy with the monolayer geometry~\cite{Margetis20}. In Sec.~\ref{sec:Dispersion}, the problem at hand is solved exactly via the Wiener-Hopf method~\cite{MGKrein1962,Gohberg1960}. 

\subsection{Symmetric and antisymmetric edge states}
\label{subsec:transf-pot}
Next, we introduce the symmetric and antisymmetric modes, which are characterized by transformed scalar potentials of the form $\varphi_1(x)\pm \varphi_2(x)$. This characterization is motivated below, being related to the concepts of the bulk optical and acoustic plasmons, respectively, on infinitely extended, translationally invariant layers~\cite{Stauber14}.

By adding and subtracting the equations of Eq.~\eqref{eq:phi1-int} (for $j=1$ and $j=2$), we find 
\begin{align}\label{eq:int-phs}
&\varphi^\pm(x)=\frac{\I\omega\mu}{k_0^2}(\sigma_0\pm\sigma_1)(\partial_x^2-q^2)\int_0^\infty \d x'\, \mathfrak K^{\pm}(x-x')\,\varphi^\pm(x')\notag\\
&-\frac{\I\omega\mu}{k_0^2}\left\{\left[(\sigma_0\pm\sigma_1)\partial_x-\I q \left(\sigma_B\pm\sigma_B'\right)\right] \mathfrak K^{\pm}(x)\,\varphi^\pm(0^+)\right.\notag\\
&\mbox{}\qquad \left. \pm\I q \sigma_2\mathfrak K^{\pm}(x)\,\varphi^\mp(0^+)\right\}~.	
\end{align}
In the above, we use the definitions
\begin{equation}\label{eq:phi-sa}
\varphi^\pm(x)=\varphi_1(x)\pm\varphi_2(x)~,\quad \mathfrak K^{\pm}(x)=\mathfrak K_\parallel(x)\pm\mathfrak K_\perp(x)~,	
\end{equation}
where $\varphi^+$ ($\varphi^-$) corresponds to the symmetric (antisymmetric) state with the corresponding kernels. Simultaneously, we will also use the notation $\phs=\varphi^+$ and $\pha=\varphi^-$ including the corresponding kernels, $\mathfrak K_{\text{S}, \text{A}}=\mathfrak K^\pm$. The two integral equations are coupled only if $\sigma_2\neq 0$.

The alert reader may notice that the right-hand side of Eq.~\eqref{eq:int-phs} may blow up at $x=0$ for the singular kernel. Despite this behavior, the potentials can be continuous across the edge for suitable values of $\omega(q)$ which allow for appropriate cancellation of the singular terms. 

As an alternative to the singular kernels, we also discuss the effect of regularized kernels. These can be constructed by replacement of $\mathfrak{K}_{\text{m}}(x)$ with $\mathfrak K_{\text{m}}(\sqrt{x^2+b_{\text{m}}^2})$ for m=S,\,A; and express edge broadening, horizontally by length $b_{\text{S}}$~\cite{Volkov88} and vertically by $b_{\text{A}}$ ($|q|b_{\text{m}}\ll 1$). 

For fully translation-invariant layers, the integration range of the integral equations for $(\phs(x),\pha(x))$ becomes the whole real axis, without any boundary terms. The resulting decoupled dispersion relations amount to the familiar bulk optical ($\phs$) and acoustic ($\pha$) plasmons. The former mode has a dispersion relation of the form $\omega^2/\sqrt{k_x^2+q^2}\simeq {\rm const.}$ via the lossless Drude model for $\sigma_0+\sigma_1$, where $(k_x,q)$ is the wave vector in the $xy$-plane~\cite{Low17}. The acoustic bulk mode has a dispersion relation of the form $(\omega/\sqrt{k_x^2+q^2})d^{-1/2} \simeq \text{const.}$~\cite{Hwang09}. However, especially for the acoustic mode, nonlocal corrections can become important~\cite{Santoro88}. In fact, the local approximation for the conductivity used here can only be applied if the sound velocity is larger than the Fermi velocity, $v_F$~\cite{StauberPRB12}.

Moreover, for the \emph{bulk modes} in the double-layer system, optical plasmons  are composed of in-phase current excitations leading to an oscillating electric dipole. These current excitations lead to transverse (in-plane) out-of-phase current excitations which give rise to an oscillating magnetic dipole. Electric and magnetic dipoles are thus collinear, which in fact defines chiral excitations, and the two moments are related via $\sigma_2$. However, the plasmonic dispersion relation is only modified by retardation effects which are proportional to both $\sigma_2$ and $v_F/c$~\cite{Lin20}. 

Due to the one-dimensional nature of edge modes, on the other hand, transverse out-of-phase fluctuations together with longitudinal in-phase fluctuations are not possible. Nevertheless, we will find a coupling between optical (electric-dipole) modes and acoustic (magnetic-dipole) modes. The electric and magnetic dipoles  are not collinear but mutually perpendicular; see Fig.~\ref{fig:Dipoles}. The coupling leads to a modified dispersion relation depending on $\sigma_2$ in the nonretarded limit.  This coupling should also modify the spin-momentum coupling which is inherent to localized nanophotonic modes~\cite{Stauber2DM19}.

\section{Dispersion relation of edge modes}
\label{sec:Dispersion}
\color{black}
In this section, we derive the dispersion relation of the edge modes in the quasi-electrostatic approach under the isotropic conductivity model of Sec.~\ref{sec:Formulation}; see Eq.~\eqref{eqs:J-sigma-E}. We use the long-range electrostatic interaction with a logarithmically singular kernel. The key idea is to reduce the system displayed in Eq.~\eqref{eq:int-phs} to a single, self-consistent scalar equation. Subsequently, we apply a variant of the Wiener-Hopf method for scalar integral equations on the half line~\cite{MGKrein1962,Masujima-book}. Some technical details of derivations are provided in Appendix~\ref{app:W-H}. Approximate formulas for the edge mode dispersion are discussed in Sec.~\ref{sec:Numerics-iso}. 
 
\subsection{Field equation and self-consistency condition}
\label{subsec:gen-int-eq}
We address the solution of Eq.~\eqref{eq:int-phs} by exploiting the property that the associated convolution integrals are decoupled. The coupling of symmetric and antisymmetric edge states occurs via the boundary (edge) terms.  

We proceed to outline the main steps. The first step is to introduce an integral equation that captures the form of Eq.~\eqref{eq:int-phs}. Consider the equation
\begin{align}\label{eq:phi-generic}
\phi(x)&=\frac{\I\omega\mu}{k_0^2}\sigma (\partial_x^2-q^2)\int_0^\infty \d x'\,\mathcal K(x-x')\,\phi(x')\notag\\
&\mbox{} \quad -\frac{\I\omega\mu}{k_0^2}[c_1 \sigma \partial_x \mathcal K(x)+ c_2  \bs \I q  \mathcal K(x)]~,\ \mbox{all}\ x~,	
\end{align}
where $c_1$, $c_2$, $\sigma$ and $\bs$ are constants. By comparison of the above equation to Eq.~\eqref{eq:int-phs}, we  identify the function $\phi$ with the potential $\phs=\varphi^+$ or $\pha=\varphi^-$ and the kernel $\mathcal K$ with $\mathfrak K_{\text{S}}=\mathfrak K_\parallel+\mathfrak K_\perp$ or $\mathfrak K_{\text{A}}=\mathfrak K_\parallel-\mathfrak K_\perp$. The parameters $\sigma$, $\bs$, $c_1$ and $c_2$ are chosen accordingly, e.g., $\phi(0^+)=c_1$. Our next step is to \emph{derive a relation among $c_1$, $c_2$, $\omega$ and $q$}, which we view as a self-consistency condition, so that the potential $\phi(x)$  is integrable and continuous~\cite{Note}. 

We apply the Fourier transform with respect to $x$. Let $\xi=k_x$ be the Fourier variable, which expresses the wave number parallel to the sheets and perpendicular to each edge. Equation~\eqref{eq:phi-generic} yields
\begin{subequations}\label{eqs:phi-generic-FT}
\begin{align}\label{eq:phi-generic-FT}
\hph_+(\xi)+\mathcal P(\xi) \hph_-(\xi)=-\frac{\I\omega\mu}{k_0^2}(\I c_1\sigma \xi+\I c_2 \bs q)\widehat{\mathcal K}(\xi)
\end{align}
for real $\xi$, where 
\begin{equation}\label{eq:P-def}
	\mathcal P(\xi)=1+\frac{\I\omega\mu\sigma}{k_0^2}	\beta(\xi)^2\widehat{\mathcal K}(\xi)~,\ \beta(\xi)=\sqrt{\xi^2+q^2}~,
\end{equation}
\end{subequations}
and $\widehat{\mathcal K}$ is the kernel Fourier transform. Bear in mind that
\begin{equation*}
	\widehat{\mathcal K}(\xi)=\frac{1}{2\beta(\xi)}\left[1\pm e^{-\beta(\xi)d}\right]~,\quad \Re\beta(\xi)>0~,
\end{equation*} 
for the symmetric ($+$) or antisymmetric ($-$) case. In the above, $\hph_{\mp}(\xi)$  is the Fourier transform of $\phi(x)$ for $x>0$ ($-$) or $x<0$ ($+$); thus,  $\hph=\hph_+ +\hph_-$. The interested reader is referred to Appendix~\ref{app:W-H} for more details. 

We should comment on the meaning of $\mathcal P(\xi)$ for given $q$. The zeros, $\xi=\xi_{\text{sp}}$, of $\mathcal P(\xi)$ that satisfy $\Re\big(\beta(\xi_{\text{sp}})\big)>0$ and $\Re(\xi_{\text{sp}})> 0$ correspond to bulk plasmonic states that propagate away from the edge, in the positive $x$-direction. This interpretation is a direct generalization of the bulk plasmons  for the monolayer configuration~\cite{Margetis20}.  

The main objective of the Wiener-Hopf method is to yield formulas for both functions $\hph_{\pm}$ from Eq.~\eqref{eq:phi-generic-FT} and the expected analytic properties of $\hph_{\pm}$. This is achieved by separating all terms in this equation into `$+$' and `$-$' functions, which are analytic in the upper and lower $\xi$-plane, respectively. This task requires the factorization of $\mathcal P(\xi)$, which means finding functions $Q_{\pm}(\xi)$ such that
\begin{equation*}
	Q(\xi)=\ln\mathcal P(\xi)=Q_+(\xi)+Q_-(\xi)\Rightarrow \mathcal P(\xi)=e^{Q_+(\xi)} e^{Q_-(\xi)}~.
\end{equation*}
In the isotropic setting, this factorization is guaranteed if $\mathcal P(\xi)$ is free of zeros in the real axis. The split functions $Q_{\pm}(\xi)$ for $Q(\xi)$ are~\cite{Masujima-book,Margetis20}
\begin{equation}\label{eq:Qpm-def}
	Q_{\pm}(\xi)=\pm \frac{1}{2\pi \I}\int_{-\infty}^\infty \d \xi'\ \frac{Q(\xi')}{\xi'-\xi}~,\ \pm\Im \xi>0~.
\end{equation}
Consequently, Eq.~\eqref{eq:phi-generic-FT} is recast to
\begin{align}\label{eq:phi-generic-FT-mod}
&e^{-Q_+(\xi)}\hph_+(\xi)+e^{Q_-(\xi)} \hph_-(\xi)\notag\\
&=-\frac{\I\omega\mu}{k_0^2}(\I c_1\sigma \xi+\I c_2 \bs q)\widehat{\mathcal K}(\xi) e^{-Q_+(\xi)}
\end{align}
for real $\xi$. Since the left-hand side is in the desired form, we need to focus on the right-hand side. The latter can be expressed as $-\I[\Lambda_+(\xi)+\Lambda_-(\xi)]$ for appropriate split functions $\Lambda_{\pm}(\xi)$, where the factor $\I$ is used for later algebraic convenience; see Appendix~\ref{app:W-H} for specifics.  

Hence, given $\Lambda_\pm$, the equation satisfied by $\hph_{\pm}$ reads
\begin{align*}
e^{-Q_+(\xi)}\hph_+(\xi)+\I \Lambda_+(\xi)=-e^{Q_-(\xi)} \hph_-(\xi)-\I \Lambda_-(\xi)
\end{align*}
for real $\xi$. Each side of this equation is analytic when continued to the respective half plane ($\Im\xi>0$ for `$+$' terms and $\Im\xi<0$ for `$-$' terms). By analytic continuation, each of these functions is equal to the same entire (everywhere-analytic) function which is a polynomial of $\xi$. The only polynomial compatible with the properties of $\mathcal K$ and $\phi$ is identically zero (see Appendix~\ref{app:W-H}).

By $\Lambda_\pm= c_1 \Lambda_{\pm}^1 + c_2 \Lambda_{\pm}^2$, we thus express $\phi(x)$ as
\begin{subequations}\label{eqs:phi-I12}
\begin{align}\label{eq:phi-I1-I2}
\phi(x)= c_1 I_1(x) + c_2 I_2(x)	
\end{align}
where
\begin{equation}\label{eq:I1-def}
I_1(x)=\frac{1}{2\pi \I}\int_{-\infty}^\infty \d \xi \ e^{\pm Q_{\pm}(\xi)}\Lambda_{\pm}^1(\xi)\,e^{\I\xi x}~,
\end{equation}
\begin{equation}\label{eq:I2-def}
I_2(x)=\frac{1}{2\pi \I}\int_{-\infty}^\infty \d \xi \ e^{\pm Q_{\pm}(\xi)}\Lambda_{\pm}^2(\xi)\,e^{\I\xi x}~,	
\end{equation}
\end{subequations}
for $\pm x<0$, using the inverse Fourier transform of $\phi$. The functions $\Lambda_\pm^{1,2}(\xi)$ are given in Eq.~\eqref{eq:app:Lambda-def} of Appendix~\ref{app:W-H}.

We can now derive a relation among $c_1$, $c_2$, $\omega$ and $q$ so that $\phi(x)$ is continuous. By the above formulas, we readily check that 
$I_1(0^+)=1$ and $I_2(0^+)=0$. Thus, we have $\varphi(0^+)=c_1$, which is consistent with Eq.~\eqref{eq:phi-generic}. We only need to study the limit values $I_1(0^-)$ and $I_2(0^-)$, in order to enforce continuity condition of $\varphi(x)$ at $x=0$. By a technical argument involving a Fourier integral, we find that $c_1 I_1(0^-)+c_2 I_2(0^-)$ diverges \emph{unless} we impose 
\begin{align}\label{eq:c1c2-generic}
c_1\sigma\left[e^{Q_-(-\I \tq)}+e^{-Q_+(\I\tq)}\right]+\I c_2\bs \sg(q)\notag\\
\times \left[e^{Q_-(-\I \tq)}-e^{-Q_+(\I \tq)}\right]=0~;	
\end{align}
see Eq.~\eqref{eq:app:c1c2-cond} of Appendix~\ref{app:W-H}. Note that $\tq= \pm q$ if $\pm\Re q>0$; thus, $\tq=|q|$ for real $q$. Equation~\eqref{eq:c1c2-generic} is the desired self-consistency condition. We can then verify that $\phi(x)$ is continuous across the edge (Appendix~\ref{app:W-H}).

\subsection{Dispersion relation unveiled}
\label{subsec:disp-reln}
Next, we invoke self-consistency condition~\eqref{eq:c1c2-generic} in order to derive the dispersion relation of 
edge modes from integral equations~\eqref{eq:int-phs}. The recipe suggested by our analysis is simple: Apply Eq.~\eqref{eq:c1c2-generic} to the integral equations for the symmetric state, $\phi=\phs$, and  the antisymmetric state, $\phi=\pha$. This procedure entails a system of linear equations for  $(\phs(0^+), \pha(0^+))$. The requirement of nonzero solutions yields the dispersion relation.

\medskip 

{\em Symmetric state.} By comparison of Eq.~\eqref{eq:phi-generic} to Eq.~\eqref{eq:int-phs} for $\phi=\varphi^+=\phs$, we set
\begin{align*}
c_1&=\varphi_{\text{S}}(0^+)~,\ \sigma= \sigma_0+\sigma_1~,\ \mathcal K=\mathfrak K_{\text{S}}=\mathfrak K_\parallel+\mathfrak K_\perp~,\\
c_2\bs &=-\left(\sigma_B+\sigma_B'\right)\varphi_{\text{S}}(0^+)+\sigma_2\varphi_{\text{A}}(0^+)	~.
\end{align*}
Thus, relation~\eqref{eq:c1c2-generic} entails 
\begin{subequations}
\begin{align}\label{eq:phisa-I}
	&\left(\sigma_{\text{S}}^- e^{Q_-^{\text{S}}(-\I\tq)}+\sigma_{\text{S}}^+ e^{-Q_+^{\text{S}}(\I \tq)}\right)\varphi_{\text{S}}(0^+) \notag \\
	&\mbox{ } +\I \sigma_2\,\sg(q) \left(e^{Q_-^{\text{S}}(-\I \tq)}-e^{-Q_+^{\text{S}}(\I \tq)}\right)\varphi_{\text{A}}(0^+)=0~,
\end{align}
where
\begin{align}
\sigma_{\text{S}}^{\pm}&=\sigma_0+\sigma_1\pm \I \sg(q)\left(\sigma_B+\sigma_B'\right)~, \label{eq:sigma-symm}
\end{align}
and $Q_\pm^{\text{S}}(\xi)$ are defined by Eq.~\eqref{eq:Qpm-def} by use of $Q=\ln\mathcal P$ and Eq.~\eqref{eq:P-def} with $\mathcal K=\mathfrak K_{\text{S}}$. Note the Fourier transform
\begin{equation*}
	\widehat{\mathfrak K}_{\text{S}}(\xi)=\widehat{\mathcal G}(\xi,0)+\widehat{\mathcal G}(\xi,d)=\frac{1+e^{-\sqrt{\xi^2+q^2}\,d}}{2\sqrt{\xi^2+q^2}}~.
\end{equation*}
\end{subequations}

\medskip

{\em Antisymmetric state.} We now set $\phi=\varphi^-=\varphi_{\text{A}}$, and
\begin{align*}
c_1&=\varphi_{\text{A}}(0^+)~,\ \sigma= \sigma_0-\sigma_1~,\ \mathcal K=\mathfrak K_{\text{A}}=\mathfrak K_\parallel-\mathfrak K_\perp~,\\
c_2\bs &= -\left(\sigma_B-\sigma_B'\right)\varphi_{\text{A}}(0^+)-\sigma_2\varphi_{\text{S}}(0^+)	~.
\end{align*}
Thus, the self-consistency condition becomes
\begin{subequations}
\begin{align}\label{eq:phisa-II}
	&\left(\sigma_{\text{A}}^- e^{Q_-^{\text{A}}(-\I\tq)}+\sigma_{\text{A}}^+ e^{-Q_+^{\text{A}}(\I \tq)}\right)\varphi_{\text{A}}(0^+) \notag \\
	&\mbox{ } -\I \sigma_2\, \sg(q)\left(e^{Q_-^{\text{A}}(-\I \tq)}-e^{-Q_+^{\text{A}}(\I \tq)}\right)\varphi_{\text{S}}(0^+)=0~,
\end{align}
where
\begin{align}\label{eq:sigma-asymm}
\sigma_{\text{A}}^{\pm}&=\sigma_0-\sigma_1\pm \I \sg(q)\left(\sigma_B-\sigma_B'\right)~.
\end{align}
The functions $Q_\pm^{\text{A}}(\xi)$ are defined by Eq.~\eqref{eq:Qpm-def} with $Q=\ln\mathcal P$. Recall Eq.~\eqref{eq:P-def} again, setting $\mathcal K=\mathfrak K_{\text{A}}$ with
\begin{equation*}
	\widehat{\mathfrak K}_{\text{A}}(\xi)=\widehat{\mathcal G}(\xi,0)-\widehat{\mathcal G}(\xi,d)=\frac{1-e^{-\sqrt{\xi^2+q^2}\,d}}{2\sqrt{\xi^2+q^2}}~.
\end{equation*}
\end{subequations}
%

\medskip

{\em Dispersion relation.} The last step of our derivation is to require that the linear system of Eqs.~\eqref{eq:phisa-I} and~\eqref{eq:phisa-II} admits solutions $(\varphi_{{\text S}}(0^+), \varphi_{\text{A}}(0^+))\neq (0,0)$. Hence, the determinant of this system should vanish, which leads to 
\begin{subequations}
\begin{align}\label{eq:disp-reln}
&\left(\sigma_{\text{S}}^- e^{\mathcal Q_{\text{S}}(q)}+\sigma_{\text{S}}^+\right)	\left(\sigma_{\text{A}}^- e^{\mathcal Q_{\text{A}}(q)}+\sigma_{\text{A}}^+\right) \notag\\
& \mbox{    }-\sigma_2^2 \left(e^{\mathcal Q_{\text{S}}(q)}-1\right)	\left(e^{\mathcal Q_{\text{A}}(q)}-1\right)=0
\end{align}
where (for m=S,\,A) 
\begin{align}\label{eq:QSA-def}
\mathcal Q_{\text{m}}(q)&=Q_+^{\text{m}}(\I q\sg(q))+	Q_-^{\text{m}}(-\I q\sg(q))\notag\\
 &=\frac{2q\sg(q)}{\pi}\int_0^\infty \d \xi\ \frac{\ln(\mathcal P_{\text{m}}(\xi))}{\xi^2+q^2}~,
\end{align}
\begin{align*}
	\mathcal P_{\text{m}}(\xi)=1+\frac{\I\omega\mu \sigma_{\text{m}}}{k_0^2}(\xi^2+q^2) \widehat{\mathfrak K}_{\text{m}}(\xi)~,
\end{align*}
\end{subequations}
and $\sigma_{\text{S}}=\sigma_0+\sigma_1$, $\sigma_{\text{A}}=\sigma_0-\sigma_1$. 
Equation~\eqref{eq:disp-reln} is the dispersion relation for edge modes under the conductivity model of Eq.~\eqref{eqs:J-sigma-E}. The two types of states are coupled through the chirality parameter, $\sigma_2$. We investigate this coupling in Sec.~\ref{sec:Numerics-iso}. The same form of dispersion relation is recovered with regularized kernels; see Appendix~\ref{app:kernel-reg}.

\section{Approximations and predictions}
\label{sec:Numerics-iso}
In this section, we discuss implications of dispersion relation~\eqref{eq:disp-reln}. We assume the long-wavelength limit according to $|qd|\ll 1$, and apply approximations to analytically capture features of the optical and acoustic edge plasmons for the isotropic conductivity model. A goal is to estimate whether the \emph{chiral coupling} between these two modes via the parameter $\sigma_2$ can be strong enough to be observed in experiments. The interested reader may directly read a summary of our results in Sec.~\ref{subsec:Summary}, skipping Sec.~\ref{subsec:decoupled}.  We also discuss the case of the neutrality point for which collective charge oscillations in the TBG are in principle not possible because of the absence of charge. In our formalism, the neutrality point is given by $\sigma_0+\sigma_1=0$, i.e., the total Drude weight  vanishes. 

By our main assumption $|q d|\ll 1$, the Fourier transforms of the singular kernels can be replaced by
\begin{subequations}\label{eqs:kernels-approx}
\begin{align}
	\widehat{\mathfrak K}_{\text{S}}(\xi)&=\frac{1+e^{-\beta(\xi) d}}{2\beta(\xi)}\simeq \frac{1}{\beta(\xi)}~,\label{eq:s-kernel-approx}\\
	\widehat{\mathfrak K}_{\text{A}}(\xi)&=\frac{1-e^{-\beta(\xi) d}}{2\beta(\xi)}\simeq \frac{d}{2}~,\quad \beta(\xi)=\sqrt{\xi^2+q^2}~,
	\label{eq:a-kernel-approx}
\end{align}
in the integrals $\mathcal Q_{\text{S,A}}$, provided $|d|\ll |\omega\mu(\sigma_0\pm\sigma_1)/k_0^2|$. 
The ensuing $\mathcal Q_{\text{A}}(q)$ is calculated in simple closed form, in contrast to $\mathcal Q_{\text{S}}(q)$. We also obtain geometric corrections for small $|qd|$; see Appendix~\ref{app:integral-iso}. The neutrality point is a special case, to be treated via a regularized kernel.

Edge broadening implies the kernel transformations
\begin{equation}\label{eq:kernel-approx-reg}
	\widehat{\mathfrak K}_{\text{S,A}}^b(\xi)= e^{-\beta(\xi)b_{\text{S,A}}}\frac{1\pm e^{-\beta(\xi)(\sqrt{b_{\text{S,A}}^2+d^2}-b_{\text{S,A}})}}{2\beta(\xi)}~,
\end{equation}
\end{subequations}
where $|q| b_\text{{S}}\ll 1$ and $|q| b_\text{{A}}\ll 1$ while $b_{\text{S}}$ and $b_{\text{A}}$ are of the same order as or larger than $d$. Approximations for $\widehat{\mathfrak K}_{\text{S,A}}^b$ can be applied accordingly; for example, see Sec.~\ref{subsec:decoupled}. 

 \subsection{Decoupled optical and acoustic edge modes}
 \label{subsec:decoupled}
 We first study a simple yet nontrivial scenario, namely, the case with  $\sigma_2=0$. Equation~\eqref{eq:disp-reln} reduces to 
 \begin{equation*}
 e^{\mathcal Q_{\text{S}}(q)}=-\frac{\sigma_{\text{S}}^+}{\sigma_{\text{S}}^-}\quad\mbox{or}\quad 
 e^{\mathcal Q_{\text{A}}(q)}=-\frac{\sigma_{\text{A}}^+}{\sigma_{\text{A}}^-}~.	
 \end{equation*}
 The parameters $\sigma_{\text{S},\text{A}}^{\pm}$ are defined by Eqs.~\eqref{eq:sigma-symm} and~\eqref{eq:sigma-asymm}. Regarding the optical plasmon (state $\phs$), the relation for $\omega(q)$ resembles the edge mode dispersion relation of a monolayer isotropic system with suitable effective $2\times 2$ conductivity matrix. This matrix has diagonal elements equal to $2(\sigma_0+ \sigma_1)=\sigma_T$ and opposite off-diagonal elements, $\sigma_{xy}=-\sigma_{yx}=2(\sigma_B+\sigma_B')$; cf.~Eq.~(40) in~\cite{Volkov88}. 
 
 We outline  approximations for the above dispersion relations. These schemes  provide some insight  into the case with a nonzero $\sigma_2$ which is discussed in Sec.~\ref{subsec:Summary}. 
 
 \medskip 
   
 \noindent {\em Acoustic edge plasmon.} We solve $e^{\mathcal Q_{\text{A}}(q)}=-\sigma_{\text{A}}^+/\sigma_{\text{A}}^-$, by employing Eq.~\eqref{eq:QSA-def} for m=A and approximation~\eqref{eq:a-kernel-approx}. The simplified integral for $\mathcal Q_{\text{A}}(q)$ equals (see Appendix~\ref{app:integral-iso})
 \begin{equation*}
 	\mathcal Q_{\text{A}}(q)\simeq \arcosh(2\eta_{\text{ac}}+1)~,\quad \eta_{\text{ac}}=\frac{\I\omega\mu(\sigma_0-\sigma_1)}{2k_0^2}q^2 d~.
 \end{equation*}
 Thus, by $\mathcal Q_{\text{A}}(q)=\ln(-\sigma_{\text{A}}^+/\sigma_{\text{A}}^-)$ we obtain 
 \begin{align*}
 \eta_{\text{ac}}&\simeq -\frac{(\sigma_{\text{A}}^++\sigma_{\text{A}}^-)^2}{4\sigma_{\text{A}}^+\sigma_{\text{A}}^-}=-\frac{(\sigma_0-\sigma_1)^2}{(\sigma_0-\sigma_1)^2+(\sigma_B-\sigma_B')^2}~.	
 \end{align*}
By using the Drude model for the counterflow conductivity, $\sigma_0-\sigma_1$, and $\sigma_B-\sigma_B'=0$, we find (for $\omega=\omega_-$) 
\begin{equation*}
	\omega_-^2\simeq \frac{D_0-D_1}{2\varepsilon} q^2 d~,
\end{equation*}
\color{black} 
where $D_j$ is the Drude weight for $\sigma_j$ ($j=0,\,1,\,2$). This mode is \emph{not observable} if $D_0-D_1<0$ (since $\omega_-^2<0$). Recall that the formalism leading to this result breaks down if the sound velocity is lower than the Fermi velocity. This dispersion relation is modified by chirality and a geometric correction due to $qd$ for $\mathcal Q_{\text{A}}(q)$ (Sec.~\ref{subsec:Summary}). If $\sigma_B-\sigma_B'\neq 0$, the acoustic edge mode becomes non-reciprocal (as expected). 

\medskip 

\noindent {\em Optical edge plasmon.} Let us focus on $e^{\mathcal Q_{\text{S}}(q)}=-\sigma_{\text{S}}^+/\sigma_{\text{S}}^-$ via Eq.~\eqref{eq:QSA-def} for m=S and formula~\eqref{eq:s-kernel-approx}. Define 
\begin{equation*}
\eta_{\text{op}}=\frac{\I\omega\mu (\sigma_0+\sigma_1)}{k_0}\frac{\tq}{k_0}~,\quad \tq=q\sg(q)~,	
\end{equation*} 
which enters $\mathcal Q_{\text{S}}(q)$.
By Eq.~\eqref{eq:QSA-def}, $\eta_{\text{op}}$ is $\mathcal O(1)$ if $-\sigma_{\text{S}}^+/\sigma_{\text{S}}^-$ is neither small nor large in magnitude, nor is it close to unity; then one has to resolve the dispersion relation numerically. In particular, for $\sigma_B+\sigma_B'=0$ one finds $\eta_{\text{op}}=\eta_{\text{op},0}\simeq -1.217$ which yields (for $\omega=\omega_+$)~\cite{Volkov88,Volkov86}
\begin{equation*}
	\frac{2\omega_+^2\varepsilon}{D_T |q|}=-\frac{1}{\eta_{\text{op},0}}\simeq 0.822~,
\end{equation*}
by use of the Drude model for $\sigma_0+\sigma_1$; $D_T=2(D_0+D_1)>0$ is the total Drude weight.
For nonzero $\sigma_B+\sigma_B'$ with $\sigma_B^{(\prime)}=-i(\omega_c/\omega)\sigma_{0(1)}$, the dispersion relation reads~\cite{Volkov88}
\begin{equation}\label{eq:disp-reln-univ}
\frac{2}{\pi}\int_0^\infty \d\xi\,\frac{\ln(\eta^\pm\tilde\beta(\xi)-1)}{\tilde\beta(\xi)^2}\simeq \ln\Biggl(\frac{1\pm s\sqrt{\eta^\pm}}{1\mp s\sqrt{\eta^\pm}}\Biggr)~,
\end{equation}
where $\tilde\beta(\xi)=\sqrt{1+\xi^2}$, $s=s(q)=\sqrt{2\omega_c^2\varepsilon/(D_T |q|)}$, $\pm\sg(q)\,\omega_c\ge 0$, and $\eta^\pm=|\eta_{\text{op}}|$ (for upper or lower sign). 

\begin{figure}[h]
\includegraphics[scale=0.18,trim=0.8in 0.3in 0.2in 0.4in]{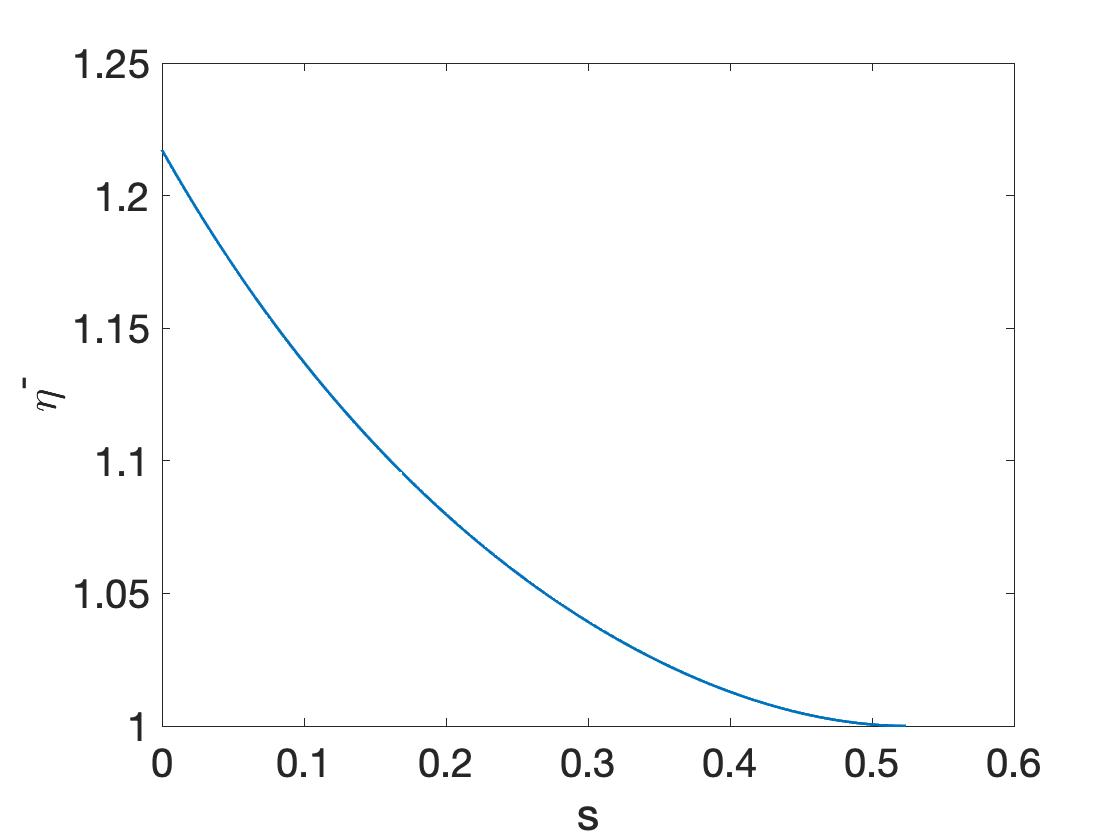}
\centering{}\caption{Universal function $|\eta_{\text{op}}|=\eta^{-}(s)$ solving Eq.~\eqref{eq:disp-reln-univ} for dispersion of edge magnetoplasmon in a single sheet with $\sg(q)\omega_c \le 0$~\cite{Volkov88} as well as for (reciprocal) optical edge plasmon in chiral TBG with $\sigma_B=\sigma_B'=0$. The range of values $0\le s< s_*\simeq 0.525$ ensures solvability of Eq.~\eqref{eq:disp-reln-univ} and implies localization of the mode. The point $(s,\eta^-)=(0,1.217)$, where $\eta^-=|\eta_{\text{op},0}|$, amounts to the known optical edge plasmon of the non-magnetic isotropic sheet~\cite{Volkov86,Volkov88}.}
\label{fig:Univ-Func}
\end{figure}

Equation~\eqref{eq:disp-reln-univ} defines the functions $\eta^{\pm}(s)$ describing non-reciprocal edge plasmons, where $\eta^{\pm}(0)=|\eta_{\text{op},0}|$. Suppose $\omega_c>0$. One mode is localized  and is dispersed according to $\eta^+(s)$ for all $q>0$ ($s>0$)~\cite{Volkov88}. For $q<0$, the other mode becomes unstable and decays into the bulk if $-|q_*|\le q<0$ ($s\ge s_*$)~\cite{Volkov88}, but is localized and dispersed via $\eta^-(s)$ for $q<-|q_*|$ ($0< s< s_*$); $|q_*|\simeq 2\omega_c^2\varepsilon/(D_T s_*^2)$, $\eta^-(s_*)=1$ and $s_*\simeq 0.525$. The universal function $\eta^-(s)$ is of particular interest and plotted in Fig.~\ref{fig:Univ-Func}. We will demonstrate that for zero out-of-plane magnetic field the chiral TBG, for $\sigma_2\neq 0$, supports optical edge plasmons that are governed by $\eta^-(s)$, and \emph{not} by $\eta^+(s)$ which would lead to a more localized mode; see Sec.~\ref{subsec:Summary}. In fact, for an intermediate range of $q$ we can show that the  chiral coupling can be interpreted as an effective magnetic field that always tends to delocalize the edge mode of a single sheet. Notably, by this correspondence we do not break reciprocity, which is usually the case with a magnetic field.

If $-\sigma_{\text{S}}^+/\sigma_{\text{S}}^-$ is either large in magnitude or close to unity, the integral for $\mathcal Q_{\text{S}}(q)$ can be computed in simple form via asymptotics (see Appendix~\ref{app:integral-iso}). In these situations, we have $|\eta_{\text{op}}|\gg 1$ or $|\eta_{\text{op}}|\ll 1$, respectively.  A small $|\sigma_{\text{S}}^+/\sigma_{\text{S}}^-|$ would imply a large negative $\Re (\mathcal Q_{\text{S}}(q))$ which is incompatible with the requisite integral.

%
%

Next, we address the case of the neutrality point ($\sigma_0+\sigma_1=0$), at which $-\sigma_{\text{S}}^+/\sigma_{\text{S}}^-$ becomes unity.  One approach is to take the \emph{limit} $\sigma_0+\sigma_1\to 0$ of the dispersion relation for the optical edge plasmon with a \emph{singular} kernel via Eq.~\eqref{eq:s-kernel-approx}. By $\eta_{\text{op}}\to 0$ with $\sigma_B+\sigma_B'\neq 0$, we obtain 
\begin{align}\label{eq:eta-neutrality}
\frac{1}{\pi}\frac{\sigma_B+\sigma_B'}{\omega_+\varepsilon}q  \left[\ln\biggl(\frac{2}{\eta_{\text{op}}}\biggr)+1\right]&\simeq -1~.
\end{align}
This equation indicates that only one edge mode may survive in this limit, since the (sign) factor $\sg(q)$ has been canceled out. This is expected, by analogy with the case of the edge magnetoplasmon~\cite{Volkov88}; however, $\sigma_B$ is now replaced by $2(\sigma_B+\sigma_B')$. 

An alternate approach relies on edge broadening. We can set $\sigma_0+\sigma_1=0$ by using the length scale $b_{\text{S}}$ in the kernel regularization; see Eq.~\eqref{eq:kernel-approx-reg}. We now proceed by two different routes. For example, we may invoke the regularized version of the relation $e^{\mathcal Q_{\text{S}}(q)}=-\sigma_{\text{S}}^+/\sigma_{\text{S}}^-$. Hence, for real $q$ and $b_{\text{S}}\gg d$ we find (see Appendix~\ref{app:integral-iso})
\begin{equation*}
\frac{1}{\pi}\frac{\sigma_B+\sigma_B'}{\omega_+\varepsilon}q K_0(|q| b_{\text{S}})\simeq -1~;\ K_0(|qb_{\text{S}}|)\simeq \ln\biggl(\frac{2}{|q| b_{\text{S}}}\biggr)-\gamma	
\end{equation*}
if $|q|b_{\text{S}}\ll 1$, where $\gamma$ is Euler's constant. Alternatively, we obtain the same relation for $q$ by resorting to integral equation~\eqref{eq:int-phs} for $\varphi_{\text{S}}=\varphi^+$ with a regularized interaction. Indeed, by setting $\sigma_0+\sigma_1=0$ and $x=0$, we have 
\begin{equation*}
	\left(1+\frac{\omega\mu(\sigma_B+\sigma_B')}{k_0} \frac{q}{k_0} \mathfrak K^b_{\text{S}}(0)\right)\phs(0)=0
\end{equation*}
where $2\pi \mathfrak K^b_{\text{S}}(0)=K_0(|q| b_{\text{S}})+K_0(|q|\sqrt{b_{\text{S}}^2+d^2})\simeq 2K_0(|q| b_{\text{S}})$, if $b_{\text{S}}$ is large compared to $d$. The value of the length $b_{\text{S}}$ is dictated by the matching of the above behavior to that of the singular kernel, as discussed in~\cite{Volkov88}.

 \subsection{Chirality effect: Summary of results}
 \label{subsec:Summary}

Next, we study the effect of nonzero $\sigma_2$. We remark that the plasmonic bulk modes of a chiral bilayer system in the retarded regime without an out-of-plane magnetic field do not depend on the chirality, and are defined by 
\begin{align*}
\frac{\omega_+^2\varepsilon}{(D_0+D_1)\tq}=1~,\qquad \frac{2\omega_-^2\varepsilon}{(D_0-D_1)q^2d}=1~.
\end{align*}
These relations are easily expressed by the parameters $\eta_{\text{op}}=\I\omega\mu(\sigma_0+\sigma_1)\tq/k_0^2$ and $\eta_{\text{ac}}=\I\omega\mu(\sigma_0-\sigma_1)q^2d/(2k_0^2)$, introduced in Sec.~\ref{subsec:decoupled}. 
In the above, $\omega_{\pm}$ denotes the frequency of the optical ($+$) or the acoustic ($-$) bulk plasmon, and $D_j$ is the Drude weight for the conductivity $\sigma_j$ ($j=0,\,1,\,2$).
The total Drude weight, $D_T=2(D_0+D_1)$, can never be negative, $D_T\geq 0$. In contrast, the sign of the magnetic Drude weight or counterflow, $D_0-D_1$, is \emph{not} fixed. For the TBG system, one finds a paramagnetic response, characterized by $D_0-D_1<0$, around the neutrality point with Fermi energy $E_F<E_{F}^t$; and a diamagnetic response, with $D_0-D_1>0$, for $E_F>E_{F}^t$. Here, $E_{F}^t$ denotes a transition energy. Thus, there is no acoustic bulk mode in the paramagnetic regime ($\omega_-^2<0$).  

In the presence of edges, the frequency squared, $\omega_+^2$, of the optical mode of a non-chiral bilayer system acquires the extra factor $1/|\eta_{\text{op},0}|$ where $\eta_{\text{op},0}\simeq -1.217$,  just as in the case of a single layer~\cite{Volkov88}. The acoustic mode remains unchanged, with a dispersion relation given by $\eta_{\text{ac}}=-1$, which for a negative counterflow Drude weight, $D_0-D_1<0$, implies that this mode is unstable and decays into the bulk (Sec.~\ref{subsec:decoupled}). \color{black}

Let us focus on the chiral bilayer system with $D_T>0$. In the end, we discuss the case with $D_T=0$. We consider $\sigma_B=\sigma_B'=0$, and real $q$. 

\medskip 

{\em Optical edge mode.}  In this case, $|\eta_{\text{op}}|$ is not large while $|\eta_{\text{ac}}|$ is typically small. If $|qd|\ll 1$, Eq.~\eqref{eq:disp-reln} gives 
\begin{equation*}
	e^{\mathcal Q_{\text{S}}(q)}\simeq -\frac{\pm\sqrt{|D_0-D_1|}D_T+2D_2^2\sqrt{\mp q^2d/(2\omega^2\varepsilon)}}{\pm\sqrt{|D_0-D_1|}D_T-2D_2^2\sqrt{\mp q^2d/(2\omega^2\varepsilon)}}~,
\end{equation*}
for $D_0-D_1>0$ (upper sign) or $D_0-D_1<0$; see Appendix~\ref{app:integral-iso} for the integral $\mathcal Q_{\text{A}}(q)\simeq 2\sqrt{\eta_{\text{ac}}}$. 

In the paramagnetic regime ($D_0-D_1<0$), we obtain 
\begin{subequations}
\begin{equation}\label{eq:disp-opt-chiralTBG}
\frac{2}{\pi}\int_0^\infty \d\xi\,\frac{\ln(|\eta_{\text{op}}|\tilde\beta(\xi)-1)}{\tilde \beta(\xi)^2}\simeq \ln\Biggl(\frac{1-s\sqrt{|\eta_{\text{op}}|}}{1+ s\sqrt{|\eta_{\text{op}}|}}\Biggr)
\end{equation}
where $\tilde\beta(\xi)=\sqrt{1+\xi^2}$, $s=s(q)=\sqrt{2|\zeta|\chi^2 |qd|}$, $\zeta=D_2^2/(D_0^2-D_1^2)$ is the effective parameter for the coupling between the optical and acoustic modes, and $\chi=D_2/D_T$ defines the chirality. Regardless of the sign of $D_2$ (if $D_2\neq 0$), Eq.~\eqref{eq:disp-opt-chiralTBG} is of the same form as the dispersion relation of a magnetoplasmon on a single sheet with $\sg(q)\omega_c<0$; cf. Eq.~\eqref{eq:disp-reln-univ}. Here, the optical edge mode dispersion is described by the universal function $|\eta_{\text{op}}|=\eta^-(s)$ where $s=s(q)$ combines the effects of geometry and chirality. Therefore, we find  
\begin{equation}\label{eq:omega-disp-opt-ch}
\frac{2\omega_+^2\varepsilon}{D_T |q|}\simeq \big\{\eta^-\big(\sqrt{2|\zeta|\chi^2 |qd|}\big)\big\}^{-1}~.
\end{equation}
\end{subequations}
Thus, $\omega_+(q)$ can be computed via Fig.~\ref{fig:Univ-Func}. In the limit of zero chirality, we recover $2\omega_+^2\varepsilon/(D_T|q|)=|\eta_{\text{op},0}|^{-1}$. 

\begin{figure}[h]
\includegraphics[scale=0.37,trim=0.6in 0.7in 0.8in 0.4in]{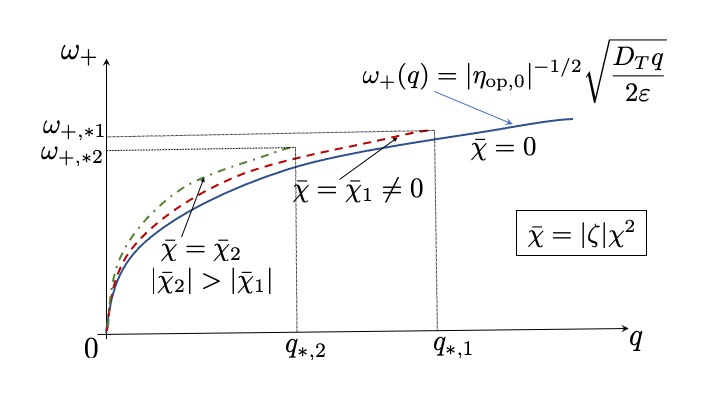}
\centering{}\caption{Schematic for effect of chirality via parameter $\bar\chi=|\zeta|\chi^2$ on frequency $\omega_+$ as a function of wave number $q$ for optical edge mode in TBG by Eq.~\eqref{eq:omega-disp-opt-ch}; $\sigma_B=\sigma_B'=0$, $D_T>0$ and $D_0-D_1<0$. Solid curve: $\bar\chi=0$. Dashed and dot-dashed curves: $\bar\chi=\bar\chi_l\neq 0$;  $q$ has a cutoff $q_{*,l}\sim \bar\chi_l^{-2}$ at frequency $\omega_{+,*l}\sim |\bar\chi_l|^{-1}$ ($l=1,\,2$). The cutoff points $(q_*,\omega_{+,*})$ obey $\eta_{\text{op}}=-1$ of the bulk mode. Near $q=0$ all dispersion curves approach the curve of zero $\bar\chi$.}
\label{fig:Schematic}
\end{figure}

If $D_2\neq 0 $ and $D_0-D_1<0$, the optical mode is localized if $0\le s(q)< s_*\simeq 0.525$ which implies $0\le |q|< |q_*|$ with cutoff wave number $|q_*|\simeq 0.138 (|\zeta|\chi^2)^{-1} d^{-1}$. The cutoff frequency is $\omega_{+,*}= \sqrt{D_T|q_*|/(2\varepsilon)}\sim |\chi|^{-1}$, which follows from $\eta_{\text{op}}=-1$ according to the bulk mode dispersion. For larger values of $|q|$, Eq.~\eqref{eq:omega-disp-opt-ch} has no admissible solution $\omega_+(q)$ and the mode decays into the bulk. Hence, in the paramagnetic regime, the smaller the parameter $|\zeta|\chi^2$ is, the wider the range of wave numbers $q$ for mode localization can be. The effect of $|\zeta|\chi^2$ on $\omega_+(q)$ is schematically shown in Fig.~\ref{fig:Schematic}. 

Our results for the optical plasmon in the chiral TBG without magnetic field suggest a correspondence of this mode to a magnetoplasmon in a single  sheet with $\sg(q)\omega_c<0$~\cite{Volkov88}; cf. Eq.~\eqref{eq:disp-reln-univ}. In the long-wavelength limit, this correspondence may not be surprising. This connection is plausible if the two systems have a common intermediate range of wave numbers $q$ supporting a localized optical mode, which can be determined through the parameter $s$ ($0\le s< s_*\simeq 0.525$).  

In order to estimate the magnetic field of the single sheet by this correspondence, we pick a value of $s$ for the TBG system according to $s=\sqrt{2|\zeta|\chi^2|q|d}$. Then, we set $s=\sqrt{2\omega_c^2\varepsilon/(D_T|q|)}$ which in turn yields the formula 
\begin{equation*}
\frac{\hbar \omega_c}{t_e}=s\sqrt{3\pi \tilde D_T \alpha_g}\sqrt{qa}~,
\end{equation*}
where $\tilde D_T t_ee^2/\hbar^2=D_T$ with $t_e=$3eV, $\alpha_g=\frac{e^2}{4\pi\varepsilon \hbar v_F}\simeq 2.2$ is the
fine-structure constant of graphene with $\hbar v_F=3 a t_e/2$, and $a=0.142$nm.
Hence, for $s=s_*$ and the typical values $qa=0.01$ and $\tilde D_T\simeq 0.01$ for the TBG~\cite{Stauber20NL}, we find an effective magnetic field $B\simeq 150$T. We observe that this effective value of $B$ is comparable to strain-induced magnetic fields in  
graphene~\cite{Levy10}. \color{black}

For $s(q)\lesssim 0.2<s_*$ the chiral effect is perturbative (with $\eta_{\text{op}}\simeq \eta_{\text{op},0}$); see Fig.~\ref{fig:Univ-Func}. The dispersion relation is (see Appendix~\ref{app:approx-disp-reln})
\begin{align}\label{eq:opt-mode-approx}
|\eta_{op,0}|\frac{2\omega_+^2\varepsilon}{D_T|q|}\simeq 1+\widetilde{C}_0\,s(q)~,
\end{align}
where $\widetilde{C}_0=\pi\sqrt{\eta_{\text{op},0}^2-1}/
[\sqrt{|\eta_{\text{op},0}|}\arccos(\eta_{\text{op},0}^{-1})]$   
for $D_0-D_1<0$. This formula explicitly shows that in the paramagnetic regime there is an undamped optical edge mode that is blue-shifted. If $0.2\lesssim s(q)<s_*\simeq 0.525$ the above formula is questionable, but the mode is still below the bulk mode and is well protected from scattering into the continuum. When $s(q)$ tends to exceed the threshold value $s_*$, however, the mode becomes delocalized. This occurs at Fermi energies well below the transition energy. 

On the other hand, for the canonical, diamagnetic regime with $D_0-D_1>0$, perturbation theory furnishes an expansion of the same form as Eq.~\eqref{eq:opt-mode-approx} albeit with $\widetilde{C}_0=-\I\pi\sqrt{\eta_{\text{op},0}^2-1}/
[\sqrt{|\eta_{\text{op},0}|}\arccos(\eta_{\text{op},0}^{-1})]$.
This suggests that the chirality leads to finite damping of the optical edge mode. \color{black} We understand this behavior by noting that the Poynting vector of the bulk plasmon forms the angle $\tan \vartheta = 2\chi q d$ with respect to the mode propagation direction $\q=q \e_y$. This direction is now fixed by the edge, which is along the $y$-axis; and the tendency of the Poynting vector to be deflected leads to dissipation of the optical edge mode. In fact, the argument involving the Poynting vector can also serve as an explanation for the tendency for  further delocalization of the edge mode in the diamagnetic as well as the paramagnetic regime. \color{black}

\medskip

{\em Acoustic edge mode.} In this case, for $D_0\pm D_1\neq 0$, $|\eta_{\text{ac}}|$ is typically of the order of unity while $|\eta_{\text{op}}|$ is large. By neglecting the geometric correction to $\mathcal Q_{\text{A}}(q)$, we derive the dispersion relation (see Appendix~\ref{app:approx-disp-reln})
\begin{equation}\label{eq:acoust-plasm-nocor}
\frac{2\omega_-^2\varepsilon}{(D_0-D_1)q^2d}\simeq \frac{1-\zeta^2}{1+2\,\text{sgn}(\zeta)s(q)^2}~.
\end{equation}
We applied the simplifying condition $|\omega_-^2 \varepsilon/(D_T q)|\ll 1$, which implies that the right-hand side of Eq.~\eqref{eq:acoust-plasm-nocor} must be kept bounded; thus, $s(q)$ must be kept small enough when $\zeta<0$ under this approximation. Interestingly, if $s(q)\ll 1$ in the paramagnetic regime ($D_0-D_1<0$, thus $\zeta<0$), for $\zeta^2>1$ the frequency $\omega_-$ of the acoustic mode becomes real with $\omega_-\sim |q|$. This property opens up the possibility of acoustic edge modes with dispersion $\omega=v_S |q|$ where the sound velocity, $v_S$, strongly depends on the chirality. Recall that $v_S$ needs to be larger than the Fermi velocity, $v_F$.  If we apply Eq.~\eqref{eq:acoust-plasm-nocor} for $s(q)\gg 1$ and $\zeta^2\gg 1$, we see that the frequency $\omega_-(q)$ is pure imaginary; thus, the mode does not seem to exist near the transition energy, $E_F^t$. 

Let us refine  the acoustic mode dispersion for $s(q)^2\ll 1$, by taking into account the geometric correction of the order of $qd$ for $\mathcal Q_{\text{A}}(q)$. We thus obtain the expression 
\begin{align}\label{eq:acoustic-mode-approx}
&\frac{2\omega_-^2\varepsilon}{(D_0-D_1)q^2d}\simeq (1-\zeta^2)\left\{1-2\,\text{sgn}(\zeta)s(q)^2\right.\notag\\
&-\frac{1}{\pi}\frac{D_0+D_1}{D_0-D_1}\zeta
\left[\ln\biggl(\frac{4}{|q| d}\biggr)+1+\I\pi\right]|q| d\notag\\
&\left.+\frac{1}{\pi}\frac{D_0+D_1}{|D_0-D_1|}\, \sqrt{\zeta^2-1}\,\arcsinh\big(\sqrt{\zeta^2-1}\big)\,|q| d\right\}~,
\end{align} \color{black}
which is a perturbative result from our analysis (see Appendix~\ref{app:approx-disp-reln}). We should also mention that in Eq.~\eqref{eq:acoustic-mode-approx} the correction term of the order of $qd$ tends to increase the real part of $\omega_-^2$, while it also causes slight damping.  

A study of the case with $D_0-D_1=0$, when $E_F$ reaches the transition energy $E_F^t$, can be carried out via the regularized kernel $\mathfrak K_{\text{A}}^b$. This study lies beyond our scope.


\medskip 

{\em The neutrality point.} 
We turn our attention to the neutrality point ($D_0+D_1=0$ for the TBG) for a few comments. The parameter $D_2$ is related to the density of states, and at the neutrality point $D_2$ vanishes due to the nature of the Dirac point, e.g., in the TBG system.  Hence, we can apply the results of Sec.~\ref{subsec:decoupled}, since the optical and acoustic modes are decoupled, including the effect of an out-of-plane magnetic field (if $\sigma_B+ \sigma_B'\neq 0$); see, e.g., Eq.~\eqref{eq:eta-neutrality}. We mention, however, that for systems (other than the TBG) with a finite density of states at the neutrality point, $D_2$ can become nonzero in this limit while $D_T\neq 0$ as well. For such systems, our results with $D_T>0$ presented in this subsection should apply.

\section{Conclusion}
\label{sec:Conclusion}
In this paper, we analytically studied  the dispersion relation of edge modes in a system of two parallel conducting layers in the nonretarded limit. Our model invokes an isotropic and spatially homogeneous conductivity tensor described by a frequency-dependent $4\times 4$ matrix, $\smatrix(\omega)$. This matrix $\smatrix$  incorporates electronic and electrostatic couplings between the two layers. Our analytical results, primarily based on the locality and isotropy of $\smatrix$, capture generic features of the edge mode dispersion in the TBG. 

We showed that chirality, which is expressed by a single parameter of the model, can cause appreciable coupling between the optical and acoustic edge modes. Regarding the optical mode, this coupling is described via a universal function in the paramagnetic regime.  We demonstrated that this mode is localized if the wave number, $q$, does not exceed a certain cutoff which decreases with increasing chirality. For an intermediate range of $q$, the chiral coupling can further be interpreted via an effective magnetic field in a corresponding single sheet. This field may become of the order of hundreds of Tesla and always tends to delocalize the edge mode.  \color{black} 

In addition, chirality opens up the possibility of observing acoustic edge modes with linear dispersion, $\omega_-(q)=v_S |q|$ where $v_S$ is the sound velocity. We believe that these results can possibly be tested in experiments.

A tool of our analysis is the Wiener-Hopf method for the coupled integral equations obeyed by scalar potentials. This approach allows us to retain the full long-range electrostatic interaction, and can be extended to an anisotropic conductivity model~\cite{Margetis21}. Our results motivate further studies in the TBG and van der Waals heterostructures, particularly the effect of the twist angle on the edge modes through a suitable conductivity tensor. 
\color{black}



\acknowledgements
The authors wish to thank G. G\'omez-Santos, T. Low, M. Luskin, and M. Maier for useful discussions. D.M. acknowledges partial support by the ARO MURI Award W911NF-14-1-0247 and the Institute for Mathematics and its Applications (NSF Grant DMS-1440471) at the University of Minnesota for several visits. The work of T.S. was supported by Spain's MINECO under Grants FIS2017-82260-P and PID2020-113164GB-I00, and by the CSIC Research Platform on Quantum Technologies PTI-001. \color{black}


\begin{appendix}

\section{Electric-field integral equations}
\label{app:int-eq}
In this appendix, we formulate a system of integral equations for the electric field
tangential to the sheets, by use of the time-harmonic Maxwell equations~\cite{Chew-book}. The formulation incorporates retardation effects; see also~\cite{DM20}. We assume that the system is described by a spatially constant, frequency-dependent conductivity tensor, $\smatrix$. This tensor is represented by a $4\times 4$ matrix. An advantage of the formalism  is the natural emergence of the condition for zero electron flux normal to each edge.

Consider the geometry of Fig.~\ref{fig:Geometry}, which consists of the flat sheets $\Sigma_1$ (at $z=0$) and $\Sigma_2$ (at $z=d$) surrounded by an isotropic and homogeneous medium of dielectric permittivity $\ed$ and magnetic permeability $\mu$. The $4$-component surface current density $\J_s=(\J_1\ \J_2)^T$ is 
\begin{equation*}
	\J_s(x,y)=
	\smatrix\cdot 
	\begin{pmatrix} \E_{\parallel}^1(x,y) \\ \E_{\parallel}^2(x,y)
	\end{pmatrix}~,\quad x>0~,
\end{equation*}
while $\J_s\equiv 0$ if $x<0$. Here, by the assumed constitutive law involving $\smatrix$, the vector $\J_j$ is the $2$-component surface current density on layer $\Sigma_j$, and $\E_\parallel^j$ is the $2$-component electric field on and tangential to sheet $\Sigma_j$. This $\E_\parallel^j$ can be defined by $\E_\parallel^j=\E-(\E\cdot \e_z)\e_z$ at $z=0$ (for $j=1$) or at $z=d$ (if $j=2$). We suppress the resulting zero transverse ($z$-) component of this vector, for algebraic convenience. The conductivity tensor is represented by a $4\times 4$ matrix of the form $\smatrix=[\smatrix_{ij}]$, where $\smatrix_{ij}$ are $\omega$-dependent $2\times 2$ matrices ($i,\,j=1,\,2$).

The volume electron current density is written as
\begin{equation*}
	\J(x,y,z)=\J_1(x,y)\,\delta(z)+\J_2(x,y)\delta(z-d)~,
\end{equation*}
where $\delta(z)$ is the Dirac delta function. This $\J$ is viewed as a $3$-component vector. We seek a system of integral equations obeyed by $\E_\parallel^j$ ($j=1,\,2$) for edge states under the following assumptions. (i) There is no current-carrying source other than $\J$. (ii) By translation invariance in $y$, the $y$-dependence of all fields is assumed to be $e^{\I q y}$. We remove this exponential by writing $\J(x,y,z)=e^{\I q y} \cJ(x,z)$, $\J_j(x,y)=e^{\I q y} \cJ_j(x)$, $\E(x,y,z)=e^{\I q y} \cE(x,z)$ and $\E_\parallel^j(x,y)=e^{\I q y} \cE_\parallel^j(x)$. The task at hand is to obtain integral equations for $\cE_\parallel^j(x)$.

The flux $\cJ(x,z)$ produces the $3$-component vector potential $\cA(x,z)$ and scalar potential $\varphi(x,z)$. In the Lorenz gauge, we have $\nabla\cdot \cA=\I k_0\varphi$ with $\nabla=(\partial_x, \I q, \partial_z)$ and 
\begin{equation*}
\cA(\r)=\mu \iint\d\rp\ G(\r-\rp)	\,\cJ(\r')~,\quad \r=(x,z)~.
\end{equation*}
Note that $\cA$ has zero $z$-component. The kernel $G(\r)$ is the appropriate Green function or propagator for the Helmholtz equation in the ambient 2D medium, viz.,
\begin{equation}\label{eq:app:Green}
G(\r)=	\frac{\I}{4} H_0^{(1)}\biggl(\sqrt{k_0^2-q^2}\,\sqrt{x^2+z^2}\biggr)~. 
\end{equation}
In the above,  $k_0^2=\omega^2\mu\ed$, $H_0^{(1)}$ is the zeroth-order modified Hankel function of the first kind, and $\Im \sqrt{k_0^2-q^2}>0$ if $q$ is real with $|q|> k_0>0$. By taking into account the structure of $\cJ$ in the bilayer system, we write
\begin{align}\label{eq:app:A-J12}
\cA(x,z)&=\mu \int_0^\infty \d x'\ G(x-x',z)\,\cJ_1(x')\notag \\
&+\mu\int_0^\infty \d x'\  G(x-x',z-d)\,\cJ_2(x')~. 
\end{align}
Note that if $\cJ_j$ are integrable, $\cA(x,z)$ is continuous.

Outside the sheets $\Sigma_1$ and $\Sigma_2$, the electric field $\cE(x,z)$ is computed by $\cE=[\I/(\omega\ed\mu)]\nabla\times\cB$ where $\cB=\nabla\times \cA$. Thus, defining $\mathcal E_x=\cE\cdot \e_x$ and $\mathcal E_y=\cE\cdot \e_y$ we obtain                          
\begin{align*}
\mathcal E_x(x,z)&=\frac{\I}{\omega\ed\mu}	\left(\I q \partial_x \mathcal A_y+q^2 \mathcal A_x-\partial_{zz}\mathcal A_x\right)~,\\
\mathcal E_y(x,z)&=-\frac{\I}{\omega\ed\mu}\left(\Delta_{xz}\mathcal A_y-\I q \partial_x \mathcal A_x\right)
\end{align*}
where $\Delta_{xz}=\partial_{xx}+\partial_{zz}$ and $\mathcal A_\ell=\e_\ell\cdot \cA$ ($\ell=x,\,y$).
A salient feature of this formalism is that $\cA(x,z)$ satisfies the homogeneous (source-free) Helmholtz equation, viz., $(\Delta_{xz}-q^2+k_0^2)\cA=0$, outside the sheets. Hence, by elimination of the derivatives $\partial_{zz} \mathcal A_x$ and $\partial_{zz} \mathcal A_y$, we express the tangential electric field  $\cE_\parallel=(\mathcal E_x, \mathcal E_y)$ as
\begin{align*}
\cE_\parallel(x,z)
=
\frac{\I}{\omega\ed\mu}
\begin{pmatrix}
\displaystyle \partial_{xx}+k_0^2 & \displaystyle \I q \partial_x \\
\displaystyle	\I q \partial_x & \displaystyle k_0^2-q^2 
\end{pmatrix}
\begin{pmatrix}
\mathcal A_x\\
\mathcal A_y	
\end{pmatrix}
~;\ z\neq 0,\,d~.
\end{align*}
By Eq.~\eqref{eq:app:A-J12}, $\cE_\parallel(x,z)$ is written explicitly in terms of the fluxes $\cJ_j$ ($j=1,\,2$). Recall that $\cJ_j=\smatrix_{j1}\cE_\parallel^1+\smatrix_{j2}\cE_\parallel^2$. Notice that $\cE_\parallel(x,z)$ is continuous since $(\mathcal A_x, \mathcal A_y)$ is.

At this stage, we can express $\cE_\parallel(x,z)$  in terms of the electric fields $\cE_\parallel^j$ on the conducting layers. If there is no charge accumulation at the edges, we may directly allow $z\to 0$ or $z\to d$ in the ensuing integral expression for $\cE_\parallel$. We expect to uncover a continuous surface current density on each sheet, including the edges. By letting $z\to 0$, we obtain a matrix equation for $\cE_\parallel^1(x)=\cE_\parallel(x,0)$; and by letting $z\to d$ we find a matrix equation for $\cE_\parallel^2(x)=\cE_\parallel(x,d)$. The resulting expression is
\begin{align}\label{eq:app:E-matrix-eqn}
&\begin{pmatrix}
\cE_\parallel^1(x)\\
\cE_\parallel^2(x)	
\end{pmatrix}
= \frac{\I \omega\mu}{k_0^2}
\begin{pmatrix}
	\underline{\mathfrak L} & 0 \\
	0         &      \underline{\mathfrak L}
\end{pmatrix} 
\int_0^\infty \d x'    \notag\\
&\times  \begin{pmatrix}
	\text{diag}(K_\parallel, K_\parallel) & \text{diag}(K_\perp, K_\perp) \\
	\text{diag}(K_\perp, K_\perp) & \text{diag}(K_\parallel, K_\parallel)
\end{pmatrix} \smatrix \begin{pmatrix}
\cE_\parallel^1(x')\\
\cE_\parallel^2(x')	
\end{pmatrix}~,
\end{align}
where $-\infty<x< \infty$, $K_\parallel=K_\parallel(x-x')$ and $K_\perp=K_\perp(x-x')$. Here, we define $K_\parallel(x)=G(x,0)$ and $K_\perp(x)=G(x,d)$, and the matrix differential operator
\begin{equation}\label{eq:app:Lop}
	\underline{\mathfrak L}=
	\begin{pmatrix}
\displaystyle \partial_{xx}+k_0^2 & \displaystyle \I q \partial_x \\
\displaystyle	\I q \partial_x & \displaystyle k_0^2-q^2 
\end{pmatrix}~.
\end{equation}
Equation~\eqref{eq:app:E-matrix-eqn} is the desired system of integral equations. 

Hence, the problem for the dispersion relation of edge states can be stated as follows. 
\emph{For given frequency $\omega$ (or wave number $q$), determine $q$ (or $\omega$) so that Eq.~\eqref{eq:app:E-matrix-eqn} has nontrivial integrable solutions $(\cE_\parallel^1,\cE_\parallel^2)$.}  The requirement of integrability of $\cE_\parallel^j(x)$  is consistent with the vanishing of the flux $\e_x\cdot \cJ_j(x)$, which is normal to the edge, as $x$ approaches the edge on each sheet~\cite{Margetis20}.


We should comment on the case when the two sheets are widely separated, as $d\to\infty$. In this limit, we should formally have $K_\perp\to 0$ while $\smatrix$ should approach a block diagonal matrix, viz.,  $\smatrix_{ij}\to 0$ for $i\neq j$ and $\smatrix_{jj}\to \smatrix_j$. Hence, Eq.~\eqref{eq:app:E-matrix-eqn} reduces to the following decoupled matrix equations, one for each layer ($j=1,\,2$): 
\begin{equation*}
	\cE_\parallel^j(x)=\frac{\I \omega\mu}{k_0^2}(\underline{\mathfrak L}\,\smatrix_j)\int_0^\infty\d x'\ K_\parallel(x-x') \cE_\parallel^j(x')~,
\end{equation*}
in agreement with the formulation for a single sheet of conductivity $\smatrix_j$~\cite{DM20}. 

\section{Quasi-electrostatic approach}
\label{app:qs}
In this appendix, we reduce the governing equations for the electric field, which are derived in Appendix~\ref{app:int-eq}, to integral equations for the scalar potential in the two  layers. The length scale over which the fields vary is small compared to the wavelength, $2\pi/k_0$, of radiation in the ambient unbounded medium~\cite{Margetis20}. This assumption implies that $|q|\gg k_0$. 

Now consider the setting (and notation) of Appendix~\ref{app:int-eq}. Application of the above scale separation implies $\cE(x,z)=-\nabla \varphi(x,z)+\I\omega\cA(x,z)\simeq -\nabla\varphi(x,z)$ where $\nabla=(\partial_x, \I q, \partial_z)$. We define 
\begin{equation*}
\varphi_1(x)=\varphi(x,0)~,\quad \varphi_2(x)=\varphi(x,d)~,
\end{equation*}
which denote the values of the scalar potential on layers $\Sigma_1$ (at $z=0$) and $\Sigma_2$ ($z=d$).
Thus, Eq.~\eqref{eq:app:E-matrix-eqn} reduces to
\begin{align}\label{eq:app:phi-int_eq1}
	&\I q 
	\begin{pmatrix}
		\varphi_1(x) \\
		\varphi_2(x)
	\end{pmatrix}\simeq\frac{\I\omega\mu}{k_0^2} 
	\begin{pmatrix}
		0 & 1 & 0 & 0 \\
		0 & 0 & 0 & 1
	\end{pmatrix}
	\begin{pmatrix}
	\underline{\mathfrak L} & 0 \\
	0         &      \underline{\mathfrak L}
\end{pmatrix} 
\int_0^\infty \d x'   \notag\\
&\times  \begin{pmatrix}
	\text{diag}(K_\parallel, K_\parallel) & \text{diag}(K_\perp, K_\perp) \\
	\text{diag}(K_\perp, K_\perp) & \text{diag}(K_\parallel, K_\parallel)
\end{pmatrix} \smatrix 
\begin{pmatrix}
\partial_{x'}\varphi_1\\
\I q \varphi_1(x') \\
\partial_{x'} \varphi_2\\	
\I q \varphi_2(x') 
\end{pmatrix} 
\end{align}
for all real $x$. By the quasi-electrostatic approach, the following approximations are also applied:
\begin{equation*}
q^2-k_0^2\simeq q^2~,\qquad \pdv[2]{}{x}+k_0^2 \simeq 	\pdv[2]{}{x}~.
\end{equation*}
In addition, Eq.~\eqref{eq:app:E-matrix-eqn} yields an analogous matrix equation in which the left-hand side involves $(\partial_x\varphi_1, \partial_x \varphi_2)$. This additional equation is redundant, since it can be obtained by differentiation of Eq.~\eqref{eq:app:phi-int_eq1} with respect to $x$. 

Equation~\eqref{eq:app:phi-int_eq1} can be recast into a simplified system of integral equations for $\varphi_1$ and $\varphi_2$ through integration by parts. By this procedure, the values of $\varphi_j(x)$ at the edge on each sheet (at $x=0^+$ for $j=1,\,2$) are singled out. Notably, the ensuing equations are compatible with the vanishing of the surface current normal to each edge, without the additional imposition of this condition. 

Hence, after some algebra, we obtain the system
\begin{align}\label{eq:app:phi-int_eq2}
\varphi_j(x)&\simeq \frac{\I\omega\mu}{k_0^2}
\begin{pmatrix}
\partial_x & \I q	
\end{pmatrix}
\sum_{i,l=1,2}\smatrix_{il}\left\{
\begin{pmatrix}
\partial_x \\
\I q	
\end{pmatrix} \right. \notag\\
&\mbox{}\times \int_0^\infty \d x'\,\mathfrak K_{ij}(x-x')\varphi_l(x')\notag\\
&\left.-
\begin{pmatrix}
1 \\
0	
\end{pmatrix}
\mathfrak K_{ij}(x)\varphi_l(0^+)\right\}~;\ j=1,\,2~,
\end{align}
for all $x$. Here, we define $\mathfrak K_{ii}=\mathfrak K_\parallel$ and $\mathfrak K_{ij}=\mathfrak K_\perp$ if $i\neq j$. The kernels $\mathfrak K_\parallel(x)$ and $\mathfrak K_\perp(x)$ come from $K_\parallel(x)=G(x,0)$ and $K_\perp(x)=G(x,d)$, respectively, by replacement of $\sqrt{k_0^2-q^2}$ with $\I q\,\text{sg}(q)=\I \tq$ (as $k_0\to 0$), where the `complex signum' function  is $\sg(q)=\pm 1$ if $\pm\Re q>0$~\cite{Margetis20}. By Eq.~\eqref{eq:app:Green} of Appendix~\ref{app:int-eq}, we find~\cite{Margetis20}
\begin{align}\label{eq:app:e-kernels}
\mathfrak K_\parallel(x)&=\frac{\I}{4} H_0^{(1)}(\I \tq |x|)=\frac{1}{2\pi}K_0(\tq |x|)~,\notag\\
\mathfrak K_\perp(x)&=\frac{1}{2\pi}K_0(\tq \sqrt{x^2+d^2})~,\ \tq=q\sg(q)~,
\end{align}
where $K_0$ is the third-kind modified Bessel function of the zeroth order. Note that $\sqrt{z} K_0(z)$ decays exponentially for large positive values of $z$. 

The problem for the dispersion relation can thus be stated as follows. \emph{For given frequency $\omega$ (or wave number $q$), determine $q$ (or $\omega$) so that system~\eqref{eq:app:phi-int_eq2} has nontrivial integrable and continuous solutions $(\varphi_1(x), \varphi_2(x))$ for all $x$}. In addition, $(\partial_x \varphi_1, \partial_x\varphi_2)$ must be integrable. The governing integral equations can be derived, alternatively, from the Poisson equation when the sole source is the surface charge induced on the sheets (Sec.~\ref{sec:Formulation}). The values $\varphi_j(0^+)$ are not a-priori known, and form part of the (nontrivial) solution for $\varphi_j(x)$ ($j=1,2$).

\section{Application of Wiener-Hopf method}
\label{app:W-H}
In this appendix, we elaborate on the solution of the system of integral equations for the potentials $\phs=\varphi_1+\varphi_2$ and $\pha=\varphi_1-\varphi_2$ under an isotropic conductivity model and singular kernels (Sec.~\ref{sec:Formulation}). We apply a variant of the Wiener-Hopf method~\cite{MGKrein1962,Masujima-book}. Regarding the application of this method to a single  conducting layer, the reader may consult~\cite{Volkov88,Margetis20}.

Our goal is to solve the system expressed by Eq.~\eqref{eq:int-phs}, for the isotropic model of Eqs.~\eqref{eq:sigma-model1}--\eqref{eq:sigma-model3}. In our analysis, a self-consistent scheme based on a single integral equation plays a central role. This equation provides a key condition which is applied to each state ($\phs$ and $\pha$) to yield the dispersion relation.  

Therefore, we focus on the equation (rewriting Eq.~\eqref{eq:phi-generic})
\begin{align}\label{eq:app:phi-generic}
\phi(x)&=\frac{\I\omega\mu}{k_0^2}\sigma (\partial_x^2-q^2)\int_0^\infty \d x'\,\mathcal K(x-x')\,\phi(x')\notag\\
&\mbox{} \quad -\frac{\I\omega\mu}{k_0^2}[c_1 \sigma \partial_x \mathcal K(x)+ c_2  \bs \I q  \mathcal K(x)]~,\ \mbox{all}\ x~,	
\end{align}
where $c_2$ is a constant and  $c_1=\phi(0^+)$. The kernel $\mathcal K$ equals $\mathfrak K_{\text{S}}$ or $\mathfrak K_{\text{A}}$ while $\phi$ is $\phs$ or $\pha$, respectively. 

Our task is to obtain a relation among $c_1$, $c_2$, $\omega$ and $q$ so that Eq.~\eqref{eq:app:phi-generic} has an integrable and continuous solution, $\phi(x)$. We repeat that $\partial_x\phi$ is also integrable (see Appendices~\ref{app:int-eq} and~\ref{app:qs}). We view the desired relation as a self-consistency condition. In particular, we need to make sure that $\phi(x)$ is continuous across the edge, at $x=0$. Once we derive the desired condition, we apply it to the integral equation system with vector variable $(\phs, \pha)$. 

Let us introduce the Fourier transform of $\phi(x)$ with independent variable $\xi=k_x$ by the formula
\begin{equation}\label{eq:app:FT-phi}
\hph(\xi)=\int_{-\infty}^\infty \d x\,\phi(x)\,e^{-\I\xi x}=\hph_+(\xi)+\hph_-(\xi)	
\end{equation}
where
\begin{equation*}
\hph_+(\xi)=\int_{-\infty}^0 \d x\,\phi(x) e^{-\I \xi x}~,\ \hph_-(\xi)= \int_0^{\infty} \d x\,\phi(x) e^{-\I \xi x}.	
\end{equation*}
Because of the integrability of $\phi(x)$, the transforms $\hph_{\pm}(\xi)$ are analytic in the upper ($+$) or lower ($-$) $\xi$-plane and $\hph_{\pm}(\xi)\to 0$ as $\xi\to \infty$.
Equation~\eqref{eq:app:phi-generic} is transformed to the Riemann-Hilbert problem expressed by~\cite{MGKrein1962} 
\begin{align*}
\hph_+(\xi)+\mathcal P(\xi) \hph_-(\xi)=-\frac{\I\omega\mu}{k_0^2}(\I c_1\sigma \xi+\I c_2 \bs q)\widehat{\mathcal K}(\xi) 
\end{align*}
for all real $\xi$, where
\begin{equation}\label{eq:app:P-def-iso}
\mathcal P(\xi)=1+\frac{\I\omega\mu\sigma}{k_0^2}	\beta(\xi)^2\widehat{\mathcal K}(\xi),\ 	\widehat{\mathcal K}(\xi)=\frac{1\pm e^{-\beta(\xi)d}}{2\beta(\xi)}
\end{equation}
and $\beta(\xi)=\sqrt{\xi^2+q^2}$; $\Re\beta(\xi)> 0$. In the above functional equation,  $\hph_{+}(\xi)$ and $\hph_-(\xi)$ are unknown. Because of their prescribed analyticity, we can determine each of these functions explicitly. The expression for $\widehat{\mathcal K}(\xi)$ in Eq.~\eqref{eq:app:P-def-iso} amounts to $\widehat{\mathfrak K}_{\text{S}}$ ($+$, upper sign) or $\widehat{\mathfrak K}_{\text{A}}$ ($-$). 

\subsection{Wiener-Hopf factorization}
\label{app:sssec:fact-meth}

Next, we apply the Wiener-Hopf method to the functional equation for $\hph_\pm(\xi)$~\cite{MGKrein1962}. We first seek functions $Q_\pm(\xi)$ analytic in the upper ($+$) or lower ($-$) $\xi$-plane such that $\mathcal P(\xi)=e^{Q_+(\xi)} e^{Q_-(\xi)}$, which amounts to
\begin{equation}\label{eq:app:Q-def}
Q(\xi)=\ln\big(\mathcal P(\xi)\big)=Q_+(\xi)+Q_-(\xi)~,
\end{equation}
assuming that $\mathcal P(\xi)$ is \emph{nonzero for all real $\xi$}. There is a technical subtlety here. To determine $Q_\pm(\xi)$ directly, we need to make sure that the logarithm of $\mathcal P(\xi)$ behaves as a single-valued function when $\xi$ takes values from $-\infty$ to $+\infty$ on the real axis. Fortunately, this property holds because $\mathcal P(\xi)$ is even. More precisely, we can assert that 
\begin{equation*}
\nu=\frac{1}{2\pi}\arg\left\{1+\frac{\I\omega\mu\sigma}{k_0^2}	\beta(\xi)^2\,\widehat{\mathcal K}(\xi)\right\}\biggl|_{\xi=-\infty}^{+\infty}=0~,	
\end{equation*}
if $\mathcal P(\xi)\neq 0$ for all real $\xi$ in the isotropic case. 
This $\nu$ is a winding number which may in principle take zero or nonzero integer values for an anisotropic model~\cite{Margetis20,Margetis21}. 

Given that $\nu=0$ for our problem, it is legitimate to apply Cauchy's integral formula to $Q(\xi)$ directly here, and write $Q_\pm(\xi)$ as~\cite{Masujima-book}
\begin{equation}\label{eq:app:Q+-}
Q_{\pm}(\xi)=\pm \frac{1}{2\pi \I}\int_{-\infty}^{\infty}\d\xi'\ \frac{Q(\xi')}{\xi'-\xi}~,	\ \pm\Im\xi>0~.
\end{equation}
We have not been able to compute these integrals exactly in simple closed form by use of known special functions.

Thus, $\hph_+$ and $\hph_-$ satisfy
\begin{align}\label{eq:app:func-eq-mod}
&e^{-Q_+(\xi)}\hph_+(\xi)+e^{Q_-(\xi)} \hph_-(\xi)\notag \\
&=-\frac{\I\omega\mu}{k_0^2}(\I c_1\sigma \xi+\I c_2 \bs q)\widehat{\mathcal K}(\xi)e^{-Q_+(\xi)}~,\ \mbox{all\ real}\ \xi.
\end{align}
In this equation, we must now completely separate the `$+$' and `$-$' parts, i.e., the functions analytic in the upper ($+$) and lower ($-$) $\xi$-plane. The objective is to find split functions $\Lambda_\pm(\xi)$ such that
\begin{equation}\label{app:eq:Lambda+-_intro}
\frac{\I\omega\mu}{k_0^2}(c_1\sigma \xi+c_2 \bs q)\widehat{\mathcal K}(\xi)e^{-Q_+(\xi)}=\Lambda_+(\xi)+\Lambda_-(\xi)~.	
\end{equation}

We proceed to calculate $\Lambda_\pm(\xi)$. Consider the identity
\begin{equation*}
	\frac{\I\omega\mu\sigma}{k_0^2}\beta(\xi)^2\widehat{\mathcal K}(\xi)\,e^{-Q_+(\xi)}=e^{Q_-(\xi)}-e^{-Q_+(\xi)}~.
\end{equation*}
Consequently, Eq.~\eqref{app:eq:Lambda+-_intro} reads
\begin{equation*}
\frac{c_1\sigma \xi+c_2 \bs q}{\sigma\beta(\xi)^2}  \left(e^{Q_-(\xi)}-e^{-Q_+(\xi)}\right)=\Lambda_+(\xi)+\Lambda_-(\xi)~.	
\end{equation*}

Now we apply the partial-fraction decomposition
\begin{equation*}
	\frac{C_{j}(\xi)}{\xi^2+q^2}
=\frac{\mathcal C_{j}^+}{\xi-\I \tq}+\frac{\mathcal C_{j}^-}{\xi+\I \tq}~,\ \tq=q\sg(q)~;\ j=1,\,2~.
\end{equation*}
Here, $C_1(\xi)=\sigma\xi$, $C_{2}(\xi)=\bs q$, $\mathcal C_1^{\pm}=\sigma/2$,  and $\mathcal C_2^+=-\I\bs\sg(q)/2=-\mathcal C_2^-$. Notice that the poles, $\pm \I \tq$,  of the above decomposition lie in the upper ($+$) or lower ($-$) half plane of complex $\xi$. Thus, Eq.~\eqref{app:eq:Lambda+-_intro} reads
\begin{align*}
	&\frac{c_1\sigma \xi+c_2 \bs q}{\sigma\beta(\xi)^2}  \left(e^{Q_-(\xi)}-e^{-Q_+(\xi)}\right)=\frac{1}{2}\left\{c_1\left(\frac{1}{\xi-\I \tq}+\frac{1}{\xi+\I \tq}\right)\right.\notag\\
	&\qquad \left. -\I c_2 \frac{\bs}{\sigma}\sg(q) \left(\frac{1}{\xi-\I \tq}-\frac{1}{\xi+\I \tq}\right)\right\}\\
	&\qquad \times \left(e^{Q_-(\xi)}-e^{-Q_+(\xi)}\right)=\Lambda_+(\xi)+\Lambda_-(\xi)~. 
\end{align*}
For later algebraic convenience, we write $\Lambda_\pm=c_1 \Lambda_\pm^1 +c_2 \Lambda_\pm^2$. The  functions $\Lambda_\pm^1$ can be computed explicitly by rearrangements of terms in the product 
\begin{align*}
\left(\frac{1}{\xi-\I \tq}\pm \frac{1}{\xi+\I \tq}\right)\left(e^{Q_-(\xi)}-e^{-Q_+(\xi)}\right)	~.
\end{align*}
We omit some details here. After some algebra, we find
\begin{align}\label{eq:app:Lambda-def}
\Lambda^1_{\pm}(\xi)&=\pm\frac{1}{2}\left(\frac{e^{Q_-(-\I \tq)}-e^{\mp Q_{\pm}(\xi)}}{\xi+ \I \tq}+\frac{e^{-Q_+(\I \tq)}-e^{\mp Q_{\pm}(\xi)}}{\xi- \I \tq}\right)~,\notag\\
\Lambda^2_{\pm}(\xi)&=\mp \frac{\bs}{\sigma} \frac{\sg(q)}{2\I}\left(\frac{e^{Q_-(-\I \tq)}-e^{\mp Q_{\pm}(\xi)}}{\xi+ \I \tq} \right. \notag\\
&\qquad \left. -\frac{e^{-Q_+(\I \tq)}-e^{\mp Q_{\pm}(\xi)}}{\xi- \I \tq}\right)~.	
\end{align}

Consequently, the equation for $\hph_\pm$ is recast into
\begin{align*}
e^{-Q_+(\xi)}\hph_+(\xi)+\I [c_1 \Lambda_+^1(\xi)+c_2 \Lambda_+^2(\xi)]\notag\\
=-e^{Q_-(\xi)}\hph_-(\xi)-\I [c_1 \Lambda_-^1(\xi)+c_2 \Lambda_-^2(\xi)]	
\end{align*}
for all real $\xi$.
The transforms $\hph_\pm(\xi)$ can be determined via the following rationale. Each side of the above equation corresponds to a function analytic in the upper ($+$) or lower ($-$) $\xi$-plane, for $\Im\xi>0$ or $\Im\xi<0$ respectively. These functions are equal to each other in the real axis. Thus, taken together these functions define an entire function, $\mathfrak E(\xi)$, i.e., a function that is analytic in the whole complex $\xi$-plane. Therefore, we have 
\begin{align*}
	e^{\mp Q_\pm(\xi)}\hph_\pm(\xi)+\I [c_1 \Lambda_\pm^1(\xi)+c_2  \Lambda_\pm^2(\xi)]=\pm\mathfrak E(\xi)~,
\end{align*}
for all real $\xi$. To determine $\hph_\pm(\xi)$ we need to find $\mathfrak E(\xi)$.

The entire function $\mathfrak E(\xi)$ can be figured out by inspection of the large-$|\xi|$ behavior of the respective expressions involving $\hph_\pm(\xi)$ in the upper or lower $\xi$-plane. At this stage, it is imperative to invoke the structure of the kernel $\mathcal K(x)$.
Since $\mathcal K(x)$ is logarithmically singular at $x=0$, which stems from the behavior of $\mathcal G(x,0)$, we have 
\begin{equation*}
	\widehat{\mathcal K}(\xi)=\mathcal O(1/\xi)\ \mbox{as}\ \xi\to\infty~.
\end{equation*}
Accordingly, we can show that
\begin{equation}\label{eq:app:Q+-:asympt1}
	e^{Q_\pm(\xi)}=\mathcal O(\sqrt{\xi})\ \mbox{as}\ \xi\to\infty~;
\end{equation}
see, e.g., Eqs.~(B.1) and (B.2) of appendix~B in~\cite{Margetis20}. We infer that $\Lambda_\pm^{1,2}(\xi)\to 0$ as $\xi\to\infty$. Furthermore, recall that $\hph_\pm(\xi)\to 0$ as $\xi\to\infty$.
Hence, we deduce that  $\mathfrak E(\xi)\to 0$ in the upper $\xi$-plane while, by a quick inspection of $e^{Q_-(\xi)}\hph_-(\xi)$, we see that $\mathfrak E(\xi)$ cannot grow as fast as $\sqrt{\xi}$ in the lower $\xi$-plane. In fact, the integrability of $\partial_x\phi(x)$ implies that $e^{Q_-(\xi)}\hph_-(\xi)\to 0$ as $\xi\to\infty$; thus, $\mathfrak E(\xi)\to 0$ as $|\xi|\to \infty$. By resorting to Liouville's theorem of complex analysis, we can prove that the only entire function that accomodates all these requirements is $\mathfrak E(\xi)=0$, for all complex $\xi$. This assertion entails 
\begin{equation*}
\hph_{\pm}(\xi)=-\I e^{\pm Q_{\pm}(\xi)}\left[c_1 \Lambda_{\pm}^1(\xi)+c_2 \Lambda_{\pm}^2(\xi)\right]~.	
\end{equation*}

By the inverse Fourier transform for $\phi(x)$, we compute 
\begin{align}\label{eq:app:phi}
\phi(x)&=c_1 I_1(x)+c_2 I_2(x)
\end{align}
where 
\begin{equation}\label{eq:app:I1}
I_1(x)=\frac{1}{2\pi \I}\int_{-\infty}^\infty \d \xi \ e^{\pm Q_{\pm}(\xi)}\Lambda_{\pm}^1(\xi)\,e^{\I\xi x}	~,
\end{equation}
\begin{equation}\label{eq:app:I2}
I_2(x)=\frac{1}{2\pi \I}\int_{-\infty}^\infty \d \xi \ e^{\pm Q_{\pm}(\xi)}\Lambda_{\pm}^2(\xi)\,e^{\I\xi x}	~,
\end{equation}
if $\pm x<0$. Recall that $\Lambda^{1,2}_\pm(\xi)$ are given by Eq.~\eqref{eq:app:Lambda-def}. 

\subsection{Self-consistency condition}
\label{app:sssec:solvab}

We proceed to relate $c_1$ and $c_2$ to the values $\phi(0^+)$ and $\phi(0^-)$. By contour integration in the $\xi$-plane, we find
\begin{equation*}
I_1(0^+)=1~,\quad I_2(0^+)=0~,	
\end{equation*}
which imply that 
\begin{equation}\label{eq:app:c1-phi}
\phi(0^+)=c_1~.	
\end{equation}
This equality trivially confirms that the $c_1$ term  comes from integration by parts in Eq.~\eqref{eq:app:phi-generic}. The values $I_{1,2}(0^+)$ are implications of the asymptotic behavior of $e^{Q_-(\xi)}$ as $\xi\to\infty$, which is intimately connected to the logarithmic singularity of the kernel, $\mathcal K(x)$, at $x=0$. 

In regard to $\phi(0^-)$, taking the limit $x\uparrow 0$ is a more delicate procedure because it leads to possibly divergent integrals~\cite{Margetis20}. By manipulation of the Fourier integral $I_1(x)$ for $x<0$, we obtain the expression
\begin{align*}
	I_1(x)&= \frac{\I\omega\mu\sigma}{2k_0^2}\left\{\left[e^{Q_-(-\I \tq)}+e^{-Q_+(\I \tq)}\right] \right.\notag\\
	 &\quad \times \frac{1}{2\pi \I}\int_{-\infty}^\infty \d\xi\ e^{\I\xi x}\,\xi e^{-Q_-(\xi)}\widehat{\mathcal K}(\xi)\notag \\
	 &+\I q \left[e^{-Q_+(\I \tq)}-e^{Q_-(-\I \tq)}\right]  \notag\\
	 &\quad \times \left.\frac{1}{2\pi \I}\int_{-\infty}^\infty \d\xi\, e^{\I\xi x} e^{-Q_-(\xi)} \widehat{\mathcal K}(\xi)\right\}~,\ x<0~.
\end{align*}
In the limit as $x\uparrow 0$, the integral of the second line behaves as $1/\sqrt{|x|}$ whereas the remaining integral approaches a finite value. The situation is different for a regularized kernel, since all corresponding integrals are absolutely convergent at $x=0$ (see Appendix~\ref{app:kernel-reg}).

In a similar vein, regarding $I_2(x)$ we have the formula
\begin{align*}
	I_2(x)&= \I \sg(q) \frac{\I\omega\mu\bs}{2k_0^2}\left\{\left[e^{Q_-(-\I \tq)}-e^{-Q_+(\I \tq)}\right] \right.\notag\\
	 &\quad \times \frac{1}{2\pi \I}\int_{-\infty}^\infty \d\xi\ e^{\I\xi x}\,\xi e^{-Q_-(\xi)}\widehat{\mathcal K}(\xi)\notag \\
	 &-\I q \left[e^{-Q_+(\I \tq)}+e^{Q_-(-\I \tq)}\right]  \notag\\
	 &\quad \times \left.\frac{1}{2\pi \I}\int_{-\infty}^\infty \d\xi\, e^{\I\xi x} e^{-Q_-(\xi)} \widehat{\mathcal K}(\xi)\right\}~,\ x<0~.
\end{align*}
The integral of the second line is the same as the respective integral for $I_1(x)$ above, and diverges as $x\uparrow 0$. 

To eliminate the overall divergence at $x=0^-$ and ensure the continuity of $\phi(x)=c_1 I_1(x)+c_2 I_2(x)$, we impose 
\begin{align}\label{eq:app:c1c2-cond}
c_1\sigma\left[e^{Q_-(-\I \tq)}+e^{-Q_+(\I\tq)}\right]+\I c_2\bs \sg(q)\notag\\
\times \left[e^{Q_-(-\I \tq)}-e^{-Q_+(\I \tq)}\right]=0~.	
\end{align}
This relation is the desired self-consistency condition.

By virtue of Eq.~\eqref{eq:app:c1c2-cond} we can directly show the continuity of $\phi$ at the edge ($x=0$). To this end, we use the remaining (convergent) integrals to obtain
\begin{align*}
	\phi(0^-)=\frac{1}{4}\left\{c_1(e^{-Q_+}-e^{Q_-})-\I c_2\sg(q) \frac{\bs}{\sigma}(e^{Q_-}+e^{-Q_+})\right\} \notag\\
	\mbox{} \times (e^{Q_+}-e^{-Q_-})
\end{align*}
where $Q_{\pm}=Q_\pm(\xi)$ are evaluated at $\xi=\pm\I \tq$. By Eqs.~\eqref{eq:app:c1c2-cond} and~\eqref{eq:app:c1-phi} we see that $\phi(0^-)=c_1=\phi(0^+)$.

\section{On the kernel regularization}
\label{app:kernel-reg}
In this appendix, we entertain the scenario that the electrostatic interaction is regularized. This means that the kernel  $\mathfrak K_{\text{S}}(x)$ or $\mathfrak K_{\text{A}}(x)$ is replaced by 
\begin{equation}\label{eq:app:K-reg}
\mathfrak K_{\text{m}}^b(x)=\mathfrak K_{\text{m}}\bigl(\sqrt{x^2+b_{\text{m}}^2}\bigr)	
\end{equation}
for m=S or A. The length $b_{\text{m}}$ should satisfy $|qb_{\text{m}}|\ll 1$, but the ratio $b_{\text{m}}/d$ is $\mathcal O(1)$ or large. The regularization for m=S is invoked in Sec.~\ref{subsec:decoupled} at the neutrality point.

We focus on Eq.~\eqref{eq:phi-generic} for $\phi=\varphi_{\text{m}}$ (m=S or A) under replacement~\eqref{eq:app:K-reg}. Hence, we solve
\begin{align}\label{eq:app:phi-generic-reg}
\phi(x)&=\frac{\I\omega\mu}{k_0^2}\sigma (\partial_x^2-q^2)\int_0^\infty \d x'\,\mathcal K^b(x-x')\,\phi(x')\notag\\
&\mbox{} \quad -\frac{\I\omega\mu}{k_0^2}[c_1 \sigma \partial_x \mathcal K^b(x)+ c_2  \bs \I q  \mathcal K^b(x)]~,
\end{align}
where now $\mathcal K^b$ is $\mathfrak K_{\text{S}}^b$ or $\mathfrak K_{\text{A}}^b$. Note that $c_1=\phi(0^+)$, consistent with the derivation of the integral equations. The Fourier transform of $\mathfrak K_{\text{m}}^b(x)$ is given by Eq.~\eqref{eq:kernel-approx-reg}. Thus, by defining $Q_\pm(\xi)$ via Eqs.~\eqref{eq:app:Q+-} and~\eqref{eq:app:Q-def}, with $\widehat{\mathcal K}$ replaced by $\widehat{\mathcal K}^b$, we can assert that 
\begin{equation}\label{eq:app:Q+-:asympt2}
Q_\pm(\xi)\to 0\ \mbox{as}\ \xi\to\infty	~,
\end{equation}
in contrast to formula~\eqref{eq:app:Q+-:asympt1} in Appendix~\ref{app:W-H}.

\subsection{Wiener-Hopf factorization process revisited}
\label{app:subsec:W-H_reg}
We start by taking the Fourier transform of Eq.~\eqref{eq:app:phi-generic-reg} with respect to $x$. The factorization process with a regularized kernel leading to formulas for $\hph_{\pm}(\xi)$ is similar to that for a logarithmically singular kernel (Appendix~\ref{app:W-H}). By inspection of the relevant formulas and use of Eq.~\eqref{eq:app:Q+-:asympt2}, we can still assert that the entire function is $\mathfrak E(\xi)=0$, for all complex $\xi$. Thus, again we find  
\begin{equation*}
\hph_{\pm}(\xi)=-\I e^{\pm Q_{\pm}(\xi)}\left[c_1 \Lambda_{\pm}^1(\xi)+c_2\Lambda_{\pm}^2(\xi)\right]~,	
\end{equation*}
where $Q_\pm$ and $\Lambda_{\pm}^{1,2}$ are defined by Eqs.~\eqref{eq:app:Q+-} and~\eqref{eq:app:Lambda-def} with Eq.~\eqref{eq:app:Q-def} (Appendix~\ref{app:W-H}) under the replacement of $\widehat{\mathcal K}$ by $\widehat{\mathcal K}^b$. Thus, the potential $\phi(x)$ is written as the linear combination $c_1 I_1(x)+c_2 I_2(x)$, Eq.~\eqref{eq:app:phi}, with the functions $I_1(x)$ and $I_2(x)$ defined by Eqs.~\eqref{eq:app:I1} and~\eqref{eq:app:I2}.

Interestingly, for $b=b_{\text{m}}\ge 0$ we compute
\begin{align*}
	I_1(0^+)&=1-\frac{1}{2} \left\{e^{-Q_+(\I \tq)}+e^{Q_-(-\I \tq)}\right\} e^{-Q_-(\infty)}~,\notag\\
	I_2(0^+)&=-\frac{\sg(q)}{2\I}\frac{\bs}{\sigma}\left\{e^{-Q_+(\I\tq)}-e^{Q_-(-\I \tq)}\right\}e^{-Q_-(\infty)}~.
\end{align*}
In particular, for $b=b_{\text{m}}> 0$ we have $Q_-(\infty)=0$ by Eq.~\eqref{eq:app:Q+-:asympt2}. On the other hand, in the case with $b=0$ (singular kernel) we see that $Q_-(\infty)=\infty$, thus recovering the values $I_1(0^+)=1$ and $I_2(0^+)=0$ of Appendix~\ref{app:W-H}.

\subsection{Self-consistency condition via regularization}
\label{app:subsec:solv_reg}

Next, we show that the relation among $c_1$, $c_2$, $\omega$ and $q$ for $b=b_{\text{m}}> 0$ is given by Eq.~\eqref{eq:app:c1c2-cond}, with $\widehat{\mathcal K}$ replaced by $\widehat{\mathcal K}^b$. The details leading to this relation are different.

First, by $\phi(0^+)=c_1 I_1(0^+)+c_2 I_2(0^+)$ we require that
\begin{align}\label{eq:app:I12-0-reg}
\phi(0^+)&=c_1\left\{1-\frac{1}{2}\left[e^{-Q_+(\I \tq)}+e^{Q_-(-\I \tq)}\right]\right\}\notag\\
  &\mbox{} -c_2 \frac{\sg(q)}{2\I}\frac{\bs}{\sigma}\left[e^{-Q_+(\I \tq)}-e^{Q_-(-\I\tq)}\right]~.	
\end{align}
Recall that $\phi(0^+)=c_1$, which implies relation~\eqref{eq:app:c1c2-cond}.

It is of interest to check whether $\phi(x)$ is continuous at the edge, viz., $\phi(0^+)=\phi(0^-)$. We claim that this continuity is satisfied without any extra condition if $b>0$. By computation of $I_{1,2}(0^-)$ for $b\ge 0$, we can write
\begin{align*}
I_1(0^-)&=1-\frac{1}{2}\left[e^{Q_-(-\I \tq)}+e^{-Q_+(\I \tq)}\right]e^{Q_+(\infty)}	~,\notag\\
I_2(0^-)&=\frac{\sg(q)}{2\I}\frac{\bs}{\sigma} \left[e^{Q_-(-\I\tq)}-e^{-Q_+(\I\tq)}\right] e^{Q_+(\infty)}~.
\end{align*}
In the special case with $b=0$ (singular kernel), we have $Q_+(\infty)=\infty$; thus, the respective Fourier integrals appear divergent, as expected (see Appendix~\ref{app:W-H}). For $b> 0$, we use Eqs.~\eqref{eq:app:Q+-:asympt2} and~\eqref{eq:app:I12-0-reg} to directly verify that
\begin{equation*}
\phi(0^-)=c_1 I_1(0^-)+c_2 I_2(0^-)=\phi(0^+)~.	
\end{equation*}
%

\section{Evaluation of integrals}
\label{app:integral-iso}
In this appendix, we compute in simple closed forms key integrals that pertain
to the dispersion relation of edge modes in the isotropic TBG system (Sec.~\ref{sec:Dispersion}), when $|q d|\ll 1$.
The analysis is needed for the theory of Sec.~\ref{sec:Numerics-iso}. 

We focus on integrals $\mathcal Q_{\text{A}}(q)$ and $\mathcal Q_{\text{S}}(q)$ of Eq.~\eqref{eq:QSA-def} via the approximations of Eqs.~\eqref{eq:s-kernel-approx} and~\eqref{eq:a-kernel-approx} for the singular kernel; or, Eq.~\eqref{eq:kernel-approx-reg} for a regularized kernel (m=S). We also derive geometric corrections for small $|qd|$.

\medskip

{\em Integral $\mathcal Q_{\text{A}}(q)$.} This case pertains to the state $\varphi_{\text{A}}$. After a change of variable, the integral $\mathcal Q_{\text{A}}(q)$ with a singular kernel under Eq.~\eqref{eq:a-kernel-approx} is written as
\begin{align*}
\mathcal Q_{\text{A}}(q)&\simeq \mathcal Q_{\text{A},0}(q)=\frac{1}{\pi}\int_{-\infty e^{-\I \arg(\tq)}}^{+\infty e^{-\I \arg(\tq)}}\d\xi\ \frac{\ln[1+\eta(1+\xi^2)]}{1+\xi^2}\\
&= \ln\eta + \frac{1}{\pi}\int_{-\infty e^{-\I \arg(\tq)}}^{+\infty e^{-\I \arg(\tq)}}\d\xi\ \frac{\ln\big[\xi+\I\sqrt{1+1/\eta}\big]}{1+\xi^2}\\
&\qquad + \frac{1}{\pi}\int_{-\infty e^{-\I \arg(\tq)}}^{+\infty e^{-\I \arg(\tq)}}\d\xi\ \frac{\ln\big[\xi-\I\sqrt{1+1/\eta}\big]}{1+\xi^2}~,
\end{align*}
where 
\begin{equation*}
\eta=\eta_{\text{ac}}=\frac{\I\omega\mu(\sigma_0-\sigma_1)}{2k_0^2}q^2d\quad (\eta\neq 0)~.
\end{equation*}
For approximation~\eqref{eq:a-kernel-approx} to make sense we must have $|\eta_{\text{ac}}|> \mathcal O((qd)^2)$.  
We compute the last two integrals by contour integration, closing the path in the upper or lower $\xi$-plane via the residue theorem. Thus, we find
\begin{equation}\label{app:eq:Qa-iso}
	\mathcal Q_{\text{A},0}(q)= \ln\big[2\eta+1+2\sqrt{\eta(\eta+1)}\big]=\arcosh(2\eta+1)~.
\end{equation}
We repeat that this leading-order result follows from kernel approximation~\eqref{eq:a-kernel-approx}, for $|qd|\ll 1$. We use the inverse hyperbolic cosine $w=\arcosh(\zeta)$ with $0\le \Im w\le \pi$. Note that $\mathcal Q_{\text{A}}(q)\simeq \mathcal Q_{\text{A},0}(q)\simeq 2\sqrt{\eta_{\text{ac}}}$ if $|qd|^2\ll |\eta_{\text{ac}}|\ll 1$.

Let us now derive a correction for $\mathcal Q_{\text{A}}(q)$ that accounts for the next-order term, of the order of $(qd)^2$ ($|qd|\ll 1$), in the expansion for $\widehat{\mathfrak K}_{\text{A}}(\xi)$. We approximate ($\tq=q\sg(q)$)
\begin{equation*}
	1-e^{-\tq d\sqrt{1+\xi^2}}\simeq \tq d \sqrt{1+\xi^2}\left[1 -\frac{1}{2}\tq d \sqrt{1+\xi^2}\right]~.
\end{equation*}
Hence, $\mathcal Q_{\text{A}}(q)$ becomes $\mathcal Q_{\text{A}}\simeq \mathcal Q_{\text{A},0}+\mathcal Q_{\text{A},1}$ where
\begin{align*}
	\mathcal Q_{\text{A},1}(q)&=\frac{2}{\pi}\int_0^{\infty e^{-\I \arg(\tq)}}\d\xi\ (1+\xi^2)^{-1}\\
	&\times \ln\Biggl\{1-\frac{1}{2}\eta\,(\tq d)\frac{(1+\xi^2)^{3/2}}{1+\eta\,(1+\xi^2)}\biggr\}~,
\end{align*}
and $\mathcal Q_{\text{A},0}(q)$ is the zeroth-order term computed above. We assume that $\eta=\eta_{\text{ac}}=\mathcal O(1)$. For small $|qd|$, the major contribution to integration in the integral for $\mathcal Q_{\text{A},1}$ comes from large $|\xi|$. After some manipulations, we write
\begin{align*}
	\mathcal Q_{\text{A},1}(q)&\simeq \frac{2}{\pi}\int_0^{\infty e^{-\I\arg(\tq)}}\d\xi\ \frac{\ln\big(1-\frac{1}{2}\tq d\xi\big)}{1+\xi^2}+\frac{1}{\pi} \tq d \\
	& \times \int_0^{\infty e^{-\I\arg(\tq)}}\d\xi\,\Biggl\{\frac{\xi}{1+\xi^2}-\eta\frac{\sqrt{1+\xi^2}}{1+\eta\,(1+\xi^2)}\biggr\}~.
\end{align*}
The first integral can be computed, via appropriate analytic continuation, from the related integral
\begin{equation*}
	\mathcal I(\epsilon)=\frac{2}{\pi}\int_0^\infty \d\xi\ \frac{\ln(1+\epsilon\xi)}{1+\xi^2}~,\quad 0<\epsilon\ll 1~.
\end{equation*}
We evaluate $\d \mathcal I(\epsilon)/\d\epsilon$ and then integrate in $\epsilon$ using $\mathcal I(0)=0$. Thus, we compute $\mathcal I(\epsilon)\simeq (2/\pi) (\epsilon\ln\epsilon^{-1}+\epsilon)$ for $0<\epsilon\ll 1$, and subsequently analytically continue the result to $\epsilon=-\tq d/2$. The integral in the second line of the formula for $\mathcal Q_{\text{A},1}(q)$ is computed by contour integration via the change of variable $\xi=\sinh t$. We find
\begin{align}\label{eq:app:QA1-form}
	\mathcal Q_{\text{A},1}(q)&\simeq -\frac{\tq d}{\pi} \Biggl\{\ln\Biggl(\frac{4}{\tq d}\Biggr)+1\notag \\
	& -(1+\eta_{\text{ac}})^{-1/2}\,\arcsinh(\eta_{\text{ac}}^{-1/2})+\I\pi\Biggr\}~,
\end{align}
where $-\pi/2\le \Im w \le \pi/2$ with $w=\arcsinh(\zeta)$. 

\medskip 

{\em Integral $\mathcal Q_{\text{S}}(q)$.} In this case, we invoke the parameter
\begin{equation*}
\eta=\eta_{\text{op}}=\frac{\I\omega\mu(\sigma_0+\sigma_1)}{k_0^2}\tq~.
\end{equation*}
The integral of interest with a singular kernel is 
\begin{align*}
\mathcal Q_{\text{S}}(q)&\simeq \mathcal Q_{\text{S},0}(q)=\frac{2}{\pi}\int_{0}^{+\infty e^{-\I \arg(\tq)}}\d\xi\ \frac{\ln\big(1+\eta\sqrt{1+\xi^2}\big)}{1+\xi^2}~,
\end{align*}
which comes from approximation~\eqref{eq:s-kernel-approx} provided $|\eta_{\text{op}}|> \mathcal O(qd)$.
We have been unable to express this integral exactly in terms of simple transcendental functions; see also~\cite{Volkov88,Margetis20}. Hence, we resort to asymptotics.

Consider the regime with $|\eta|\gg 1$. By writing 
\begin{equation*}
\ln(1+\eta\sqrt{1+\xi^2})= \ln\eta+\ln(\eta^{-1}+\sqrt{1+\xi^2})
\end{equation*}
 and neglecting $\eta^{-1}$ in the logarithm, we approximate
\begin{align}\label{app:eq:Qs-iso-large}
	\mathcal Q_{\text{S},0}(q)&\simeq \frac{2}{\pi}\int_0^{+\infty e^{-\I \arg(\tq)}}\d\xi\ \frac{\ln\eta+\ln\big(\sqrt{1+\xi^2}\big)}{1+\xi^2}\notag\\
	 &= \ln(2\eta)\quad \mbox{if}\ \ |\eta|\gg 1~. 
\end{align}

We now examine the regime with $|\eta|\ll 1$. The respective computation is not essentially different from that of the integral $\mathcal I(\epsilon)$ regarding the correction $\mathcal Q_{\text{A},1}(q)$ above; see also~\cite{Margetis20}. We thus obtain the asymptotic formula
\begin{equation}\label{app:eq:Qs-iso-small}
\mathcal Q_{\text{S},0}(q)\simeq \frac{2}{\pi}\eta \big[\ln(2/\eta)+1\big]\quad \mbox{if}\ \ |\eta|\ll 1~.	
\end{equation}

Next, we derive a correction term for $\mathcal Q_{\text{S}}(q)$ that takes into account the next-order term in the expansion for $\widehat{\mathfrak K}_{\text{S}}(\xi)$ in powers of $qd$ ($|qd|\ll 1$).  We approximate
\begin{equation*}
	1+e^{-\tq d\sqrt{1+\xi^2}}\simeq 2-\tq d \sqrt{1+\xi^2}
\end{equation*}
and write $\mathcal Q_{\text{S}}\simeq \mathcal Q_{\text{S},0}+\mathcal Q_{\text{S},1}$, where $\mathcal Q_{\text{S},0}$ is the zeroth-order term (discussed above) and
\begin{align*}
	\mathcal Q_{\text{S},1}(q)&=\frac{2}{\pi} \int_0^{\infty e^{-\I\arg(\tq)}}\d\xi\ (1+\xi^2)^{-1}\\
	&\times \ln\Biggl\{1-\frac{\eta\tq d}{2}\frac{1+\xi^2}{1+\eta\sqrt{1+\xi^2}}\Biggr\}~. 
\end{align*}
We assume that $\eta=\eta_{\text{op}}=\mathcal O(1)$. For $|qd|\ll 1$, the major contribution to integration arises from large $|\xi|$. After some manipulations, we obtain the expansion
\begin{align*}
\mathcal Q_{\text{S},1}(q)&\simeq -\frac{\tq d}{\pi}\Biggl\{\ln\Biggl(\frac{4}{\tq d}\Biggr)+1-\frac{1}{\eta}\frac{\arccos(\eta^{-1})}{\sqrt{1-\eta^{-2}}}+\I\pi\Biggr\}	
\end{align*}
where $0\le \Re w\le \pi$ with $w=\arccos(\zeta)$.

Next, we turn our attention to the effect of regularization regarding the symmetric state. The integral reads
\begin{align*}
\mathcal Q_{\text{S}}(q)&\simeq \frac{2}{\pi}\int_{0}^{+\infty e^{-\I \arg(\tq)}}\d\xi\ \frac{\ln\big(1+\eta e^{-\tq b\sqrt{1+\xi^2}}\sqrt{1+\xi^2}\big)}{1+\xi^2}
\end{align*}
for $b=b_{\text{S}}\gg d$, by Eq.~\eqref{eq:kernel-approx-reg} for $\widehat{\mathfrak K}_{\text{S}}$.
We evaluate this integral for $|\eta|\ll 1$. By Taylor expanding in powers of $\eta$ the numerator of the integrand, we obtain
\begin{align}\label{eq:app:Qs-regul-int}
\mathcal Q_{\text{S}}(q)&\simeq \frac{2\eta}{\pi}\int_{0}^{+\infty e^{-\I \arg(\tq)}}\d\xi\ \frac{e^{-\tq b\sqrt{1+\xi^2}}}{\sqrt{1+\xi^2}}=\frac{2\eta}{\pi} K_0(\tq b)~,
\end{align}
via the change of variable $\xi=\sinh t$. This result can be simplified for $|qb|\ll 1$ by use of $K_0(\tq b)\simeq \ln(2/(\tq b))-\gamma$ where $\gamma=0.577215\ldots$ is Euler's constant.  Equation~\eqref{eq:app:Qs-regul-int} can be easily modified if $b=\mathcal O(d)$ by use of an additional term with $K_0(\tq \sqrt{b^2+d^2})$. The result applies to the optical plasmon at the neutrality point (Sec.~\ref{subsec:decoupled}).

Let us now perturb the regularized integral $\mathcal Q_{\text{S}
}(q)$ via  $\eta=\eta_0(1+\epsilon)$, or $\tq=\tq_0(1+\epsilon)$, with $|\epsilon|\ll 1$. By expanding
\begin{align*}
&\ln\big[1+\eta e^{-\tq b \tilde\beta(\xi)}\tilde\beta(\xi)\big]= \ln\big[1+\eta_0 e^{-\tq_0 b \tilde\beta(\xi)}\tilde\beta(\xi)\big] \\
& \ +\epsilon\eta_0 \frac{e^{-\tq_0 b\tilde\beta(\xi)}\tilde\beta(\xi)}{1+\eta_0 e^{-\tq_0 b\tilde\beta(\xi)}\tilde\beta(\xi)}+\mathcal O(\epsilon\tq_0 b)+\mathcal O(\epsilon^2)
\end{align*}
where $\tilde\beta(\xi)=\sqrt{1+\xi^2}$, we obtain $\mathcal Q_{\text{S}}(q)=\mathcal Q_{\text{S}}^0+\eta_0\epsilon \mathcal I_1(\tq_0b)+o(\epsilon)$; $\mathcal Q_{\text{S}}^0$ equals $\mathcal Q_{\text{S}}$ at $q=q_0$ and the term $o(\epsilon)$ approaches $0$ faster than $\epsilon$. Hence, let us compute 
\begin{equation*}
\mathcal I_1(\tq_0 b)=\frac{2}{\pi}\int_0^{+\infty}\d\xi \ \frac{1}{\tilde\beta(\xi)} \frac{e^{-\tq_0 b\tilde\beta(\xi)}}{1+\eta_0 e^{-\tq_0 b\tilde\beta(\xi)}\tilde\beta(\xi)}~,
\end{equation*}
for real $\eta_0 <-1$ and  positive $\tq_0 b$ with $\tq_0 b\ll 1$. The result will be applied for other complex values of $\eta_0$ and $\tq_0 b$ with $|\tq_0 b|\ll 1$ by analytic continuation. We observe that $\mathcal I_1(x)$ is a continuous function of $x$ and converges absolutely for all real $x$, while $\mathcal I_1(x)-\mathcal I_1(0)$ vanishes as $\mathcal O(x)$ in the limit $x\to 0$. Thus, the kernel regularization is unnecessary for this calculation. Therefore, setting $b=0$ we focus on the integral
\begin{equation*}
	\mathcal I_1(\tq b)\simeq \mathcal I_1(0)=\frac{1}{\pi}\int_{-\infty}^{+\infty}\d\xi\ \frac{1}{\sqrt{1+\xi^2}}\frac{1}{1+\eta_0\sqrt{1+\xi^2}}~,
\end{equation*}
which is computed  via the change of variable $\xi=\sinh t$. This integral is conveniently written as 
\begin{equation*}
	\mathcal I_1(0)=\frac{1}{\pi \eta_0}\lim_{\delta_1\downarrow 0}\int_{-\infty}^{+\infty}\d t\ \frac{e^{-\delta_1 t}}{\cosh t+(\eta_0)^{-1}}~.
\end{equation*}
The $\delta_1$-dependent integral is evaluated for $0<\delta_1<1$ by contour integration. By applying the residue theorem to a contour of a large rectangle, we finally obtain
\begin{equation}\label{eq:app:I1-pert}
\mathcal I_1(0)=\frac{2}{\pi\eta_0} \frac{\arccos(1/\eta_0)}{\sqrt{1-(1/\eta_0)^2}}=-\frac{2}{\pi} \frac{\arcosh(1/\eta_0)}{\sqrt{1-\eta_0^2}}~.	
\end{equation}
The last expression serves the analytic continuation of $\mathcal I_1(0)$ to complex $\eta_0$ with $|\eta_0|<1$. We used the identity $\arccos(\xi)=-\I\, \arcosh(\xi)$ where the branch of $w=\arcosh(\xi)$ is defined so that $0\le \Im w\le \pi$. Recall that
$\arcosh(\xi)=\ln\big(\xi+\sqrt{\xi^2-1}\big)$. Thus, Eq.~\eqref{eq:app:I1-pert} yields a finite value of $\mathcal I_1(0)$ at $\eta_0=1$ but diverges as $\mathcal O\big(1/\sqrt{1+\eta_0}\big)$ if $\eta_0\to -1$. These findings are consistent with the definition of integral $\mathcal I_1(0)$. 
%
%

\section{Approximations of chiral dispersion}
\label{app:approx-disp-reln}
In this appendix, we outline perturbative calculations capturing the effect of chirality on the dispersion of the optical and acoustic edge plasmons away from the neutrality point (see Sec.~\ref{subsec:Summary}). We use the singular kernel with $|qd|\ll 1$. In our calculations, we set $\sigma_B=\sigma_B'=0$.

\medskip

{\em Optical edge mode.} Dispersion relation~\eqref{eq:disp-reln} gives
\begin{equation}\label{eq:app:disp-opt}
\mathcal Q_{\text{S}}(q)=\ln\Biggl[-\frac{(\sigma_0^2-\sigma_1^2+\sigma_2^2)e^{\mathcal Q_{\text{A}}(q)}+\sigma_0^2-\sigma_1^2 -\sigma_2^2}{(\sigma_0^2-\sigma_1^2 -\sigma_2^2)e^{\mathcal Q_{\text{A}}(q)}+\sigma_0^2-\sigma_1^2 +\sigma_2^2}\Biggr]~.	
\end{equation}
For $\sigma_2=0$, this relation reduces to $\mathcal Q_{\text{S}}(q)=\ln(-1)=-\I\pi$, which is approximately satisfied for 
\begin{equation*}
\eta_{\text{op}}=\frac{\I\omega\mu(\sigma_0+\sigma_1)}{k_0^2} \tilde q\simeq -(0.822)^{-1}\simeq -1.217=\eta_{\text{op},0}~,
\end{equation*}
to the leading order in $qd$. 
This solution also results from Eq.~\eqref{eq:app:disp-opt} by setting $\mathcal Q_{\text{A}}(q)=0$ with arbitrary $\sigma_2$.
We will carry out perturbations in $qd$, not in $\sigma_2$, by treating $\mathcal Q_{\text{A}}(q)$ as small in Eq.~\eqref{eq:app:disp-opt}. We consider $\eta_{\text{op}}$ as an $\mathcal O(1)$ quantity, with unperturbed value $\eta_{\text{op},0}$. Note that we could expand $\mathcal Q_{\text{S}}(q)\simeq \mathcal Q_{\text{S},0}(q)+\mathcal Q_{\text{S},1}(q)$ by taking into account the geometric correction due to the expansion of $\widehat{\mathfrak K}_{\text{S}}(\xi)$ in powers of $qd$ (Appendix~\ref{app:integral-iso}). However, this additional complication is not needed  here. 

We point out the following types of contributions in Eq.~\eqref{eq:app:disp-opt}: (i) The chirality effect (terms proportional to $\sigma_2^2$); (ii) the $qd$-dependent geometric correction term from $\mathcal Q_{\text{S}}$; and (iii) the effect of $\mathcal Q_{\text{A}}$,  the interaction with the acoustic plasmon. The $O[qd \ln((qd)^{-1})]$ contribution of item (ii) is subdominant to terms from item (iii).

Let us explain the approximation for $e^{\mathcal Q_{\text{A}}}$. Recall that the integral for $\mathcal Q_{\text{A}}$ is controlled only by the parameter
\begin{equation*}
\eta_{\text{ac}}=\frac{i\omega\mu(\sigma_0-\sigma_1)}{2k_0^2} q^2d~.
\end{equation*}
Since we take $\eta_{\text{op}}=\mathcal O(1)$, we see that
\begin{equation*}
|\eta_{\text{ac}}|=\frac{1}{2} |\eta_{\text{op}}| \left|\frac{\sigma_0-\sigma_1}{\sigma_0+\sigma_1}\right| |qd|\ll 1~,
\end{equation*}
if $|(\sigma_0-\sigma_1)/(\sigma_0+\sigma_1)|$ is small compared to $|qd|^{-1}$. This condition is plausible away from the neutrality point. Hence, we use approximation~\eqref{app:eq:Qa-iso} of Appendix~\ref{app:integral-iso}, which implies that $e^{\mathcal Q_{\text{A}}}\simeq 1+\mathcal Q_{\text{A}}\simeq 1+2\sqrt{\eta_{\text{ac}}}$ where $\Im\sqrt{\eta_{\text{ac}}}\ge 0$. Thus, the correction term due to the influence of the acoustic plasmon is $\mathcal O(\sqrt{qd})$. This effect dominates over the geometric correction for $\mathcal Q_{\text{S}}$.

There is one more step that we should take.  In Eq.~\eqref{eq:app:disp-opt}, the left-hand side needs to be perturbed around $\eta_{\text{op},0}$ in order to balance the interaction with the acoustic plasmon from $\mathcal Q_{\text{A}}$. For this purpose, we expand $\eta_{\text{op}}=\eta_{\text{op},0}+\eta_1$ ($|\eta_1|\ll |\eta_{\text{op}, 0}|$), and use perturbative formula~\eqref{eq:app:I1-pert} of Appendix~\ref{app:integral-iso} in order to determine $\eta_1$. By combining the above steps and using the Drude weights $D_j$ ($j=0,\,1,\,2$), after some algebra we obtain Eq.~\eqref{eq:opt-mode-approx} for $D_0-D_1<0$, and its counterpart for $D_0-D_1>0$. The perturbative formula holds if
$\big|\widetilde{C}_0 s(q)|\ll 1$.	

\medskip

{\em Acoustic edge mode.} In this case, we write dispersion relation~\eqref{eq:disp-reln} as
\begin{equation}\label{eq:app:disp-acoust}
\mathcal Q_{\text{A}}(q)=\ln\Biggl(-\frac{\sigma_0^2-\sigma_1^2+\sigma_2^2+(\sigma_0^2-\sigma_1^2 -\sigma_2^2)e^{-\mathcal Q_{\text{S}}(q)}}{\sigma_0^2-\sigma_1^2 -\sigma_2^2+(\sigma_0^2-\sigma_1^2 +\sigma_2^2)e^{-\mathcal Q_{\text{S}}(q)}}\Biggr)~.	
\end{equation}
If $\sigma_2=0$, this equation reduces to $\mathcal Q_{\text{A}}(q)=\ln(-1)$, which for $\mathcal Q_{\text{A}}\simeq \mathcal Q_{\text{A},0}=\arcosh(2\eta_{\text{ac}}+1)$ entails $\eta_{\text{ac}}=\I\omega\mu(\sigma_0-\sigma_1)q^2d/(2k_0^2)\simeq -1$.

More generally, our scheme for resolving Eq.~\eqref{eq:app:disp-acoust} with nonzero $\sigma_2$ can be outlined as follows. We expand $\mathcal Q_{\text{A}}(q)\simeq \mathcal Q_{\text{A},0}(q)+\mathcal Q_{\text{A},1}(q)$ for $\eta_{\text{ac}}=\mathcal O(1)$ and $|\eta_{\text{op}}|\gg 1$, as discussed in Appendix~\ref{app:integral-iso}. The term $\mathcal Q_{\text{A},1}$, where $|\mathcal Q_{\text{A},1}|\ll |\mathcal Q_{\text{A},0}|,$ is the geometric correction accounting for the expansion of $\widehat{\mathfrak K}_{\text{A}}(\xi)$ in powers of $qd$. We also expand the right-hand side of Eq.~\eqref{eq:app:disp-acoust} for $|e^{-\mathcal Q_{\text{S}}}|\ll 1$. 

Let us briefly explain the approximation associated with  $e^{-\mathcal Q_{\text{S}}}$. Recall that the integral for $\mathcal Q_{\text{S}}$ is controlled by the parameter $\eta_{\text{op}}=\I\omega\mu(\sigma_0+\sigma_1)\tq /k_0^2$.
For the acoustic plasmon, we consider $\eta_{\text{ac}}=\mathcal O(1)$; thus, we have
\begin{equation*}
|\eta_{op}|=2|\eta_{ac}| \left|\frac{\sigma_0+\sigma_1}{\sigma_0-\sigma_1}\right| \frac{1}{|qd|}\gg 1~,
\end{equation*}
if $|(\sigma_0-\sigma_1)/(\sigma_0+\sigma_1)|$ is small compared to $|qd|^{-1}$. Thus, $1/\eta_{\text{op}}$ can be of the order of $qd$, and we can use  
$\mathcal Q_{\text{S}}(q)\simeq \ln(2\eta_{\text{op}})$ by Eq.~\eqref{app:eq:Qs-iso-large} of  Appendix~\ref{app:integral-iso}.

By manipulating dispersion relation~\eqref{eq:app:disp-acoust} accordingly, after some algebra we find
\begin{align}
\frac{1}{\eta_{\text{ac}}}&\simeq -1+\frac{\sigma_2^4}{(\sigma_0^2-\sigma_1^2)^2}-\frac{2\sigma_2^4}{(\sigma_0^2-\sigma_1^2)^2} \frac{1}{\eta_{\text{op}}}\notag\\
& -\left[1-\frac{\sigma_2^4}{(\sigma_0^2-\sigma_1^2)^2}\right]\frac{\sigma_2^2}{\sigma_0^2-\sigma_1^2}\mathcal Q_{\text{A},1}(q)~, \label{eq:app:eta-ac}
	\end{align}
	where the correction term $\mathcal Q_{\text{A},1}(q)$ is given by Eq.~\eqref{eq:app:QA1-form}.
Note  that $|\sigma_2^2/(\sigma_0^2-\sigma_1^2)|$ has \emph{not} been treated as small compared to unity. The small parameter in our scheme is actually $|qd|$. However, for our scheme to hold formally, in the last equation each of the last two terms (proportional to $1/\eta_{\text{op}}$ and $\mathcal Q_{\text{A},1}$) should be treated as much smaller in magnitude than the sum of the first two terms on the right-hand side. This means that $|\sigma_2^2/(\sigma_0^2-\sigma_1^2)|$ must be much smaller than $(1/|qd|)\{\ln(4/|qd|)\}^{-1}$. The manipulation of Eq.~\eqref{eq:app:eta-ac} with retainment of $\mathcal Q_{\text{A},1}(q)$ furnishes Eq.~\eqref{eq:acoustic-mode-approx} if $s(q)^2\ll 1$.  On the other hand, the neglect of $\mathcal Q_{\text{A},1}$ in Eq.~\eqref{eq:app:eta-ac} yields Eq.~\eqref{eq:acoust-plasm-nocor}.



\end{appendix}

\bibliography{TBG-MS} 

\begin{thebibliography}{89}
\bibitem [{Not()}]{Note}%
  \BibitemOpen
  \href@noop {} {}\bibinfo {note} {In addition, the electric field at $z=0,\,d$
  normal to the edge must be integrable. The precise characterization of the
  function space for the solution lies beyond our scope.}\BibitemShut {Stop}%
\end{thebibliography}
\end{document}